\documentclass[prc,aps,floatfix,groupedaddress,amsmath,amssymb]{revtex4}
\usepackage{graphicx} 
\usepackage{epsfig,ulem} 
\usepackage{dcolumn}% Align table columns on decimal point 
\usepackage{bm}% bold math
\usepackage{color}

\newcommand{\IM}{I}

\newcommand{\lV}{\mathbf{l}} % level Vector
\newcommand{\lVE}{l} % level Vector Element
\newcommand{\FV}{\mathbf{F}} % Feeding Vector
\newcommand{\FVE}{F} % Feeding Vector Element\
\newcommand{\BM}{\mathcal{B}} % Branching Matrix
\newcommand{\BME}{\mathcal{B}} % Branching Matrix Element
\newcommand{\FM}{\mathcal{F}} % Feeding Matrix
\newcommand{\FME}{\mathcal{F}} % Feeding Matrix Element

\newcommand{\tV}{\mathbf{t}} % transition Vector
\newcommand{\tVE}{t} % transition Vecor Element
\newcommand{\PV}{\mathbf{P}} % transition Probability Vector 
\newcommand{\PVE}{P} % transition Probability Vector Element
\newcommand{\AM}{\mathcal{A}} % Adjacency Matrix
\newcommand{\AME}{\mathcal{A}} % Adjacency Matrix Element
\newcommand{\PM}{\mathcal{P}} % transition Probability Matrix 
\newcommand{\PME}{\mathcal{P}} % transition Probability Matrix Element

\newcommand{\CV}{\mathbf{T}} % transition Cascade Vector
\newcommand{\CVE}{T} % transition Cascade Vector Element
\newcommand{\CM}{\mathcal{T}} % transition Cascade Matrix
\newcommand{\CME}{\mathcal{T}} % transition Cascade Matrix Element

\newcommand{\eS}{S}  % "elementary spectrum" : explicit gate condition {g1.g2. ... .gn}
\newcommand{\cS}{\mathcal{S}} % "combined spectrum" : optional gate {g1+g2+...+gN}_m
\newcommand{\cSs}{\mathcal{S}^s} % "spiked spectrum" (represents overlapping event sets)

\newcommand{\mE}{\mathcal{E}}
\newcommand{\mG}{\mathcal{G}}

\newcommand{\mP}{\mathcal{P}}

\newcommand{\ra}{\rightarrow}

\newcommand{\be}{\begin{eqnarray}}
\newcommand{\ee}{\end{eqnarray}}
\newcommand{\een}{\nonumber \end{eqnarray}}
\newcommand{\nn}{\nonumber}

\begin{document}

\title{Gamma-ray intensities in multi-gated spectra}

\author{Camille Ducoin$^1$, Guillaume Maquart$^1$, Olivier St\'{e}zowski$^1$}
\affiliation{$^1$ Univ Lyon, Universit\'e Lyon 1, CNRS/IN2P3, IPN-Lyon, F-69622, Villeurbanne, France}

\begin{abstract}
The level structure of nuclei offers a large amount and variety of information
to improve our knowledge of the strong interaction and of mesoscopic quantum systems.
Gamma spectroscopy is a powerful tool to perform such studies:
modern gamma multi-detectors present increasing performances in terms of sensitivity and efficiency,
allowing to extend ever more our 
%access to weakly populated nuclear levels.
ability to observe and characterize abundant nuclear states.
For instance, the high-spin part of level schemes often reflects intriguing nuclear shape phenomena:
this behaviour is unveiled by high-fold experimental data analysed through multi-coincidence spectra, 
in which long deexcitation cascades become observable.
Determining the intensity of newly discovered transitions is important 
to characterize the nuclear structure and formation mechanism of the emitting levels. 
However, it is not trivial to relate the apparent intensity observed in multi-gated spectra
to the actual transition intensity.
In this work, we introduce the basis of a formalism affiliated with graph theory:
we have obtained analytic expressions from which data-analysis methods can eventually be derived 
to recover this link in a rigorous way.
\end{abstract}

\maketitle

%\begin{keyword}
%Nuclear gamma-ray spectroscopy \sep 
%Cascade emission \sep
%Transition intensities \sep
%Multi-coincidence spectra \sep
%Computer data analysis \sep
%Logic and set theory \sep
%Combinatorics \sep
%Graph theory
%\end{keyword}

\section{Introduction}

Gamma spectroscopy is one of the most important experimental techniques allowing to characterize the quantum structure of atomic nuclei.
%[ Indeed, when a nucleus is produced in an excited state, its deexcitation process often involves gamma emission,
%and can involve cascades of many gamma photons emitted in a very short time (typically of the order of $10^{-9}$ seconds) : 
%due to the time resolution of detection systems, these photons are observed simultaneously.
%Their observable properties carry fondamental information about the position and nature of the transitions between excited levels,
%which has to be exctracted thanks to minutious data analysis. ]
Gamma spectra produced under selection criteria that impose coincidence relations between the photon emissions
are of particular importance. 
First, setting different coincidence conditions and observing the resulting presence or absence of gamma rays 
allows to construct the level scheme of the nucleus.
Furthermore, coincidence conditions have a selective role that is crucial
if the studied nucleus is only one of the possible exit channels of the production reaction
(e.g. fusion-evaporation, fission...),
and if we want to observe low-intensity gamma rays.
This consideration is especially relevant in now-a-days experiments aiming at the knowledge
of exotic nuclei (e.g. neutron-rich nuclei), and exotic states in nuclei (e.g. high-spin states),
both having low production cross-section.
%[ : back-ground suppression is crucial to extract significant data.]
In this context, progress in gamma-detector resolution and efficiency
has been very important in the last decades, 
and is the leading criterion for the development of new detector arrays
such as AGATA~\cite{AGATA} and GRETA~\cite{GRETA}.
%It is now possible to work with high-fold data, and access previously unobserved regions of nuclear structure
% thanks to the study of multi-gated spectra.
Working with high-fold data, multi-gated spectra allow to observe 
nuclear structure regions that would be otherwise concealed by the background.

With increasing amount and complexity of experimental data, 
%the analysis to extract valuable information is getting more and more time-consuming.For this reason, 
efforts have been dedicated to establish automated procedures
to construct level schemes on the basis of coincidence data
(see e.g.~\cite{Haslip94,Adam97,Wilson97,Jansson11,Demand11}).
Although important steps have been taken, 
these works generally conclude that 
% although a clever computer codes can help and perform part of the work,
human intervention is still crucial to obtain valid level schemes when realistic data are employed.
%and jugement on typical defaults such as fake coincidences or missing transitions are needed.
Among the cited papers, 
of specific importance for the present study is the work of
Demand et al.~\cite{Demand11}, where the relation between nuclear level scheme and graph theory is explored.
Although our goal is different, since we are mainly focused on characterizing a new transition appearing in a previously known level scheme,
the framework of graph theory has proven very useful in the treatment of our problem.
Many textbooks exist on this mathematical formalism that has wide-spread applications;
we only indicate here one of the classic references, by Bondy and Murty~\cite{Bondy76}.

Besides level-scheme solving, many other works have been dedicated to the improvement of gamma-ray data analysis. 
The most practical ones concern software developments that offer to the user an optimized environment 
to obtain and analyze gated spectra, in relation with nucleus level scheme, such as the famous toolkit
Radware~\cite{Radford95}.
Concerning the issue of intensity measurements,
we can cite works on $\gamma-\gamma$ coincidence matrices~\cite{Chemaly94},
a method focused on the effect of angular correlations~\cite{Zahn09},
and studies to quantify coincidence-summing effects, especially the analytic approach presented in~\cite{Jutier07}.
The work of Beausang et al.~\cite{Beausang95}
calls attention on the bias on intensity measurements due to the spiking effect in gated spectra,
which is particularly relevant for the present study.

The present work is focused on the issue of relating gamma-ray intensities observed in multi-gated spectra
to the corresponding absolute emission probabilities.
This relation is non-trivial, especially in the case of gating conditions
where different combinations of photons in coincidence are allowed to select an event.
It is important to establish this relation in a rigorous way 
in order to obtain accurate values of the emission probabilities,
which contain valuable information to characterize the nuclear structure
and also to study reaction mechanism in the light of level feeding.
Starting from the graph-theory-inspired framework established by 
Demand et al.~\cite{Demand11},
we have developed a formalism and analytic expressions to establish this relation on a well-controlled basis.
The present article is dedicated to the presentation of this formalism:
for clarity, this is done using simplifying assumptions that place this study in a very idealized framework
(in substance, every transition yields a photon that is fully detected).
However, this first version will be used as a sound basis for further developments, 
and we plan next to make this formalism applicable to the analysis of real experimental data.

The present article is organized as follows.
Section~\ref{Sec:GraphTheory} 
presents the level-space and transition-space treatment of nuclear structure,
in the framework of graph theory.
Section~\ref{Sec:PresentationFormalisme}
is dedicated to the detailed description of the formalism we have derived.
It includes the demonstration of the analytical relations we have obtained
to express the gated intensities in terms of the absolute emission probabilities 
and of the transition probability matrix deduced from branching ratios.
An example of application to a schematic level scheme is presented in Section~\ref{Sec:AppliSchematicLS}.
A summary and plan for future developments are given in Section~\ref{Sec:SumOutlook}.

\section{Nucleus deexcitation as an application of graph theory}
\label{Sec:GraphTheory}

The quantum states of an atomic nucleus are linked by a network of possible transitions, 
whose probabilities are determined by the properties of the physical interaction and the quantum numbers associated with the different states.
This is one of the numerous situations that can be modelled by a mathematical object called "graph".
A graph $G$ is defined as a triple $(\textsf{V},\textsf{E},\Psi)$, where $\textsf{V}$ is a set of vertices, $\textsf{E}$ a set of edges (links between vertices),
and $\Psi$ a relationship associating each edge with a pair of vertices.
The usual representation of nuclear structure is a level scheme: 
it that can be seen as the representation of a graph
for which the elements of $\textsf{V}$ are the quantum states, 
and the elements of $\textsf{E}$ are the existing transitions. 
In addition to identifying the levels associated with a given transition, 
the relationship $\Psi$ can carry some information about the probability of this transition:
in this case, $G$ is called a weighted graph.
If transitions occur in response to an excitation, they can go towards either higher or lower energy states, 
and their probabilities depend on the properties of the excitation source.
However, we will focus on the study of nuclear deexcitation cascades 
following the formation of an excited nucleus:
in this case, transitions occur only towards lower energy levels 
and their probabilities are determined by the branching ratios.
Nuclear branching ratios only depend on the structure of the nucleus under study,
and they are abundantly documented in databases such as ENSDF \cite{ENSDF}.
Each transition can happen only in one direction (from higher to lower energy),
which means that we have a directed graph 
%(also called in abbreviated way a digraph).
(also called digraph).
To summarize, the level scheme that describes the different energy states of a nucleus 
and the transitions that can occur during its deexcitation 
is modelled mathematically by a weighted digraph.

Concerning the deexcitation cascades, each one defines a directed path,
which is a sequence of distinct vertices linked by specific directed edges.
In the present work, we will consider only simple graphs: 
namely, there is no more than one edge between two vertices.
In physical terms, this means that we only consider the existence of a transition
from one energy level to another,
and we do not distinguish different kinds of transitions between these two levels.
In future work, we will also consider different kinds of transitions 
in order to distinguish for instance between gamma transitions and electronic conversions:
in this case, several edges can link two vertices, 
and the graph is no longer a simple graph.

\begin{figure}
\includegraphics[scale=0.42]{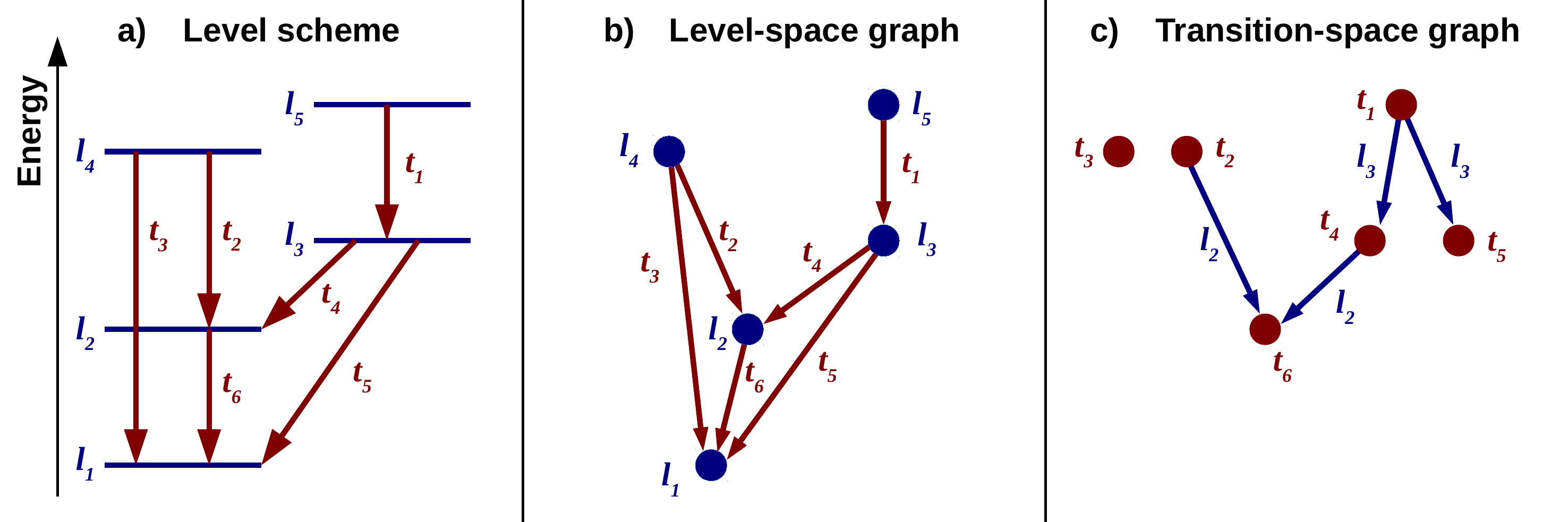}
\caption{(Color online) A simple illustration of nuclear structure representations:
(a) usual level-scheme representation, 
(b) representation of a level-space graph,
(c) representation of a transition-space graph.
Levels are denoted by $l_i$ (ordered by increasing energy),
transitions are denoted by $\tVE_i$ (ordered by decreasing energy of the emitting level).
In the graph representations, vertices are represented as dots and directed edges as arrows.
%In level space, the graph edges are weighted by the transition probabilities $P_{i}$.
%In transition space, the graph edges are weighted by the level feeding $F_i$.
}
\label{Fig:graph-theory}
\end{figure}

\subsection{Level space}

Level schemes are the usual representation of the structure of a nucleus,
and they are focused on the description of nuclear state properties. 
This point of view is called the level-space representation \cite{Demand11}.
As stated above, in terms of graph theory, nuclear levels are vertices and transitions are edges:
this is illustrated by Figure~\ref{Fig:graph-theory}.
We can notice that usually, each excited level is able to decay following a cascade down to the ground state (GS),
although nucleus disintegration may also occur before reaching the GS.
Assuming that there is always a decay branch that reaches the GS, 
we obtain a connected graph: it cannot be separated in two non-communicating sets of vertices.

Let us specify that, in the present work, the level space is limited to the discrete part of the spectrum: 
the continuum is not explicitly treated.
In this approach, in order to describe the deexcitation of a nucleus formed in a given reaction,
the following information is needed:
\begin{itemize}
\item List of possibly involved nuclear levels: vector $\lV=\{\lVE_1,...,\lVE_{D_l}\}$. 
The number of levels $D_l$ gives the dimension of the level space. 
Note that this list can vary depending on the way the excited nucleus is produced.
\item Primary feeding: vector $\FV^{(1)}=\{\FVE^{(1)}_1,...,\FVE^{(1)}_{D_l}\}$. 
Each component $\FVE^{(1)}_i$ gives the probability that level $\lVE_i$ is the first discrete level to be populated in the decay cascade.
This quantity, again, highly depends on the way the nucleus is produced.
It corresponds to either a direct feeding at the time of nucleus formation, 
or a decay from the continuum part of the spectrum.
\item Branching ratios: matrix $\BM$ of dimension $D_l \times D_l$.
One element $\BME_{ij}$ gives the probability that level $\lVE_i$ decays directly to level $\lVE_j$.
Since $\BME_{ij}$ does not depend on the way level $\lVE_i$ was formed,
these elements only depend on the nucleus itself, and can be found in nuclear databases.
Note that, in terms of graph theory, the branching matrix is the so-called adjacency matrix describing the connexions
between the vertices of a graph.
\end{itemize}

In this approach, the probability of a given transition $\tVE_x=\lVE_i \ra \lVE_j$ is given by:
$P_{x} = \FVE_i \BME_{ij}$,
where $\FVE_i$ is the total feeding of level $\lVE_i$,
i.e. the probability that level $\lVE_i$ is populated during the deexcitation.
This can be expressed as a function of the primary feeding vector $\FV^{(1)}$ 
and the branching matrix $\BM$.
Indeed, the branching matrix $\BM$ can be used to determine a secondary feeding matrix $\FM$,
where the element $\FME_{ij}$ gives the probability that level $\lVE_j$ is populated 
if level $\lVE_i$ has been populated before, with an arbitrary number of steps inbetween.
This corresponds to the following relation,
adapted from the derivation presented by Demand et al.~\cite{Demand11}:
\be 
\FME_{ij} &=& \left(\sum_{n=1}^\infty \BM^n\right)_{ij} = \left([\IM-\BM]^{-1}-\IM\right)_{ij} \nn\\
\FM &=& \sum_{n=1}^\infty \BM^n = [\IM-\BM]^{-1}-\IM \nn
\ee
where $\IM$ is the identity matrix.
For the element $\FME_{ij}$, each term of the summation over $n$ expresses the probability that $\lVE_j$
is reached $n$ steps after $\lVE_i$. 
For instance, $\BME^{2}_{ij}=\sum_k \BME_{ik}\BME_{kj}$
is the probability that $\lVE_i$ decays to $\lVE_j$ in two steps, and so on.
The second part of the equality just results from the well-known Taylor development of $[\IM-\BM]^{-1}$.
From the secondary feeding matrix $\FM$,
we obtain the secondary feeding vector $\FV^{(2)}$,
which gives the probability for each level to be populated by the decay of any other discrete level.
For each level $\lVE_i$, we have:
\be 
\FVE^{(2)}_i &=& \sum_k \FVE^{(1)}_k \FME_{ki} \nn
\ee
and the total feeding is simply given by $\FV_i=\FV^{(1)}_i+\FV^{(2)}_i$.
We can thus obtain the occurrence probability $P_x$ of each transition $\tVE_x$
using input on level-space quantities $\FV^{(1)}$ (primary feeding)
and $\BM$ (branching-ratio matrix).
However, for our purpose, we also need to express the probability of a transition
under the condition that other specific transitions (gates) occur in the same deexcitation cascade.
To treat this problem, it is more straightforward to adopt a different point of view, 
the transition-centered description of the deexcitation.

\subsection{Transition space}

As pointed out in the work of Demand et al. \cite{Demand11},
although level space offers the most natural representation of nuclear properties, 
it can be more useful in the framework of experimental data analysis to switch to a transition-centered representation.
Indeed, transitions are the experimental observables from which the level scheme has to be deduced.
The set of observed transitions is then the natural starting point 
in the search for an automated level scheme construction procedure.
Our purpose is different, 
%since it concerns a better knowledge of new levels
%to be added to a known level scheme.
since it aims at adding further knowledge to a partially known level scheme. 
However, also in our case, quantities associated with the transition space are the relevant input
needed to determine what we want: an expression of the gamma-ray intensities measured in multi-gated spectra.
In contrast with the approach of Demand et al.
where the transition space contains only the observed transitions, 
we consider here that it contains all the transitions occurring during the deexcitation process:
we make the simplifying assumption that every transition is detected,
leaving for later work further refinement of the formalism.

In the transition-space approach, in terms of graph theory, 
transitions are vertices.
The edges are links between transitions: namely, a level (in our approach),
or possibly a group of levels linked by unobserved transitions 
(in the observable-based approach developed by Demand et al.). 
Note however that, even in our case, the situation is not symmetrical to the one in the level space: 
there is no one-to-one correspondance between the set of edges and the set of levels, 
since one level can be associated with several edges (one for each of its decay modes).
This can be seen in Figure~\ref{Fig:graph-theory}, 
where we can also notice that the transition-space graph is not necessarily connected.
In order to describe the deexcitation process, the following information is needed:
\begin{itemize}
\item List of possibly involved transitions (dependent on the nucleus formation mechanism): transition vector 
$\tV=\{\tVE_1,...,\tVE_{D_t}\}$, where $D_t$ is the dimension of the considered transition space.
\item Transition probabilities: vector $\PV=\{\PVE_1,...,\PVE_{D_t}\}$, 
%giving for each transition the probability that it occurs during a deexcitation cascade
giving the probability of each transition to occur during a deexcitation cascade.
\item Adjacency matrix $\AM$, where the element $\AME_{ij}$ gives the probability that the transition $\tVE_i$ 
is immediately followed by the transition $\tVE_j$. 
\end{itemize}
Note that the above transition-space quantities
can be easily deduced from level-space input (level vector $\lV$, primary feeding vector $\FV^{(1)}$, branching matrix $\BM$):
\begin{itemize}
\item $\tV$ is obtained by listing all possible transitions from one level to the other, 
using $\lV$ and $\BM$;
\item as shown in the previous subsection, $\PV$ is deduced from $\FV^{(1)}$ and $\BM$;
\item $\AM$ is closely related to $\BM$.
Let us call $\lVE_{x,e}^{}$ the emitting level of a transition $\tVE_x$
and $\lVE_{x,r}^{}$ its receiving level:
an element $\AME_{ij}$ is non-zero only if 
%the arrival level of $\tVE_i$ is identical to the departure level of $\tVE_j$;
%in this case, it is given by the branching ratio between the two arrival levels.
$\lVE_{i,r}^{}=\lVE_{j,e}^{}$,
and in this case it is equal the branching ratio 
of the decay mode from $\lVE_{j,e}^{}$ to $\lVE_{j,r}^{}$.
\end{itemize}
We can remark that conversely, the level-space fundamental quantities ($\lV$, $\FV^{(1)}$, and $\BM$)
could be deduced from the transition-space ones ($\tV$, $\PV$, and $\AM$),
if each transition $\tVE_x$ is associated with identified emitting and receiving levels;
if not, we have to face the difficulties of level-scheme reconstruction.
We will not address this subject.
%but presently we will not make use of this property.

Let us now introduce a transition-space quantity that occupies a central place in the formalism we are developing:
the transition probability matrix $\PM$. 
The relation between $\AM$ and $\PM$ 
is analogous to the one obtained in level space 
between $\BM$ (branching matrix) and $\FM$ (secondary feeding matrix).
Namely, an element $\PM_{ij}$ gives the probability that transition $\tVE_j$ occurs
if transition $\tVE_i$ has occurred before, with an arbitrary number of steps inbetween.
This corresponds to the relation presented in Ref.~\cite{Demand11}:
\be 
\label{Eq:MatriceProba}
\PM &=& \sum_{n=1}^\infty \AM^n = [\IM-\AM]^{-1}-\IM
\ee

\section{Formalism to calculate gamma-ray intensities in gated spectra}
\label{Sec:PresentationFormalisme}

A gate condition selects events for which a given set of gamma rays are emitted in coincidence. 
The chosen gate condition has a direct impact on the presence and intensity of each gamma ray in the resulting spectrum. In this section, we develop a formalism that allows one to calculate the apparent intensity of any ray emitted during a deexcitation cascade, depending on the kind of gate condition that has been applied. We will distinguish the relatively simple case of a gate condition of type "and" from the more complex situation occurring when a gate condition of type "or" is applied.

\subsection{Simplifying hypotheses and external input}

In order to introduce the formalism, we assume some simplifying hypotheses 
(keeping for later work the generalization to more realistic cases):
\begin{itemize}
\item The nucleus emits pure gamma cascades until the ground state (no electron conversion, no decay of excited states to another nucleus by nucleon emission or $\beta$ disintegration)
\item We ignore the problem of degeneracy, which has to be considered if transitions taking place in different parts of the level scheme lead to similar gamma emissions
\item The gamma detection is ideally performed, with 100\% absolute photopeak efficiency 
%(every emitted gamma ray is detected with the correct energy).
(every emitted gamma ray is fully detected).
\end{itemize}

Furthermore, the feeding of the entry states (primary feeding, which depends on the reaction mechanism) is given as an input.

\subsection{Selected definitions}

To formalize the situation, we will need to use some specific terminology and notations.
We introduce here the most fundamental ones in order to 
settle the frame of the following discussion.

\subsubsection{Gate conditions}
\label{sssection-def-gate-conditions}

A gate condition is based on the detection of specific gamma rays, called \textit{gates}.
For each event, a gate is said to be \textit{open} when the corresponding ray is detected, 
and \textit{closed} if it is not.
The list of $N$ gates involved in the expression of a given condition will be written:
$L=\{g_1,...,g_{N}\}$, where $g_k$ identifies an individual gamma ray used as a gate. 
Depending on the way these gates are involved, we can distinguish different kinds of conditions.
For the present study, we need to define three kinds:
\begin{itemize}
\item \textit{Positive explicit gate conditions} (type "and"): all the gates of the list $L$ have to be open.
Such condition will be denoted by $G=\{g_1\cdot ... \cdot g_N\}$,
called a positive explicit condition of order $N$.
\item \textit{Exclusive explicit gate conditions}: 
each gate of the list $L$ is \textit{specified} to be either open or closed. 
This concept is particularly useful for the treatment of gate conditions of type "or", as we will see later.
The term "explicit" means that each gate of $L$ has a specified status (open or closed), 
"exclusive" means that some of them are required to be closed.
An exclusive explicit condition of order $(n,\bar{n})$ involves $n$ open gates and $\bar{n}$ closed gates,
with $n+\bar{n}=N$.
The list $L$ is then decomposed in two sublists: 
$L^{(o)}=\{g^{(o)}_1,..., g^{(o)}_n\}$ containing the open gates,
and $L^{(c)}=\{{g}^{(c)}_1, ..., {g}^{(c)}_{\bar{n}}\}$ containing the closed gates.
%These two sublists are used to identify the involved gates, 
%and classify them according to their basic status in the exclusive condition.
As we will see in Section~\ref{ssection-devel-positive-spectra},
exclusive conditions can be treated by performing developments in terms of positive conditions,
with terms involving gates from the sublist $L^{(c)}$
whose status changes from "closed" to "open".
For this reason, to express an exclusive condition,
we specify both the gate identification 
(position within the sublist $L^{(o)}$ or $L^{(c)}$)
and its status (open: $g$, or closed: $\bar{g}$). 
This leads to the notation:
$G=\{g^{(o)}_1\cdot ... \cdot g^{(o)}_n \cdot \bar{g}^{(c)}_1\cdot ... \cdot \bar{g}^{(c)}_{\bar{n}}\}$.
Although this notation may now look redundant,
% in the original expression of the exclusive condition,
it will be useful in future developments: see for instance Eq.~(\ref{Eq:spectrum-develop-step2a}).
\item \textit{Optional gate conditions} (type "or"): a minimal number of gates from the list $L$ have to be open. If $m$ is this minimal number, any event for which at least $m$ gates of $L$ are open is selected (whatever the status of the remaining gates). 
Such condition is denoted by $\mG=\{g_1+...+g_N\}_m$.
\end{itemize}
Optional gate conditions will be studied in detail in the following.
For the treatment of this case, it will be useful to consider the various explicit gate conditions 
that can be defined using sublists of $L=\{g_1,...,g_N\}$:
\begin{itemize}
\item $G_\alpha(n,L)$ denotes a positive explicit condition of order $n\leq N$:
it involves a list $L^{(\alpha)}=\{g^{(\alpha)}_1,...,g^{(\alpha)}_n\}$ that is a sublist of $L$.
For a given order $n$, the number of possible combinations of $n$ gates picked from the list $L$
is given by the well-known binomial coefficient $C^N_n=N!/[n!(N-n)!]$.
The $\alpha$ index, which identifies the different combinations, then takes the values $1\leq \alpha \leq C^N_n$. 
\item $G_\beta(n,\bar{n},L)$ denotes an exclusive explicit condition of order $(n,\bar{n})$, 
with $n+\bar{n} \leq N$: 
it involves the lists $L^{(\beta,o)}=\{g^{(\beta,o)}_1,...,g^{(\beta,o)}_n\}$ 
and $L^{(\beta,c)}=\{g^{(\beta,c)}_1,...,g^{(\beta,c)}_{\bar{n}}\}$ that are sublists of $L$.
For a given order $(n,\bar{n})$, 
there are $C^N_{n+\bar{n}}$ combinations of $n+\bar{n}$ specified gates picked from the list $L$,
and $C^{n+\bar{n}}_n$ combinations of $n$ open gates picked from the $n+\bar{n}$ specified gates.
The $\beta$ index then takes the values $1 \leq \beta \leq C^N_{n+\bar{n}}\times C^{n+\bar{n}}_n$.
\end{itemize}

\subsubsection{Associated sets of events}

Experimentally, an event corresponds to the formation of an excited nucleus and the following deexcitation cascade. In the dataset, it is identified as a list of energy deposits localized at different places in the detection system. Event reconstruction from individual deposits is a first step for data analysis, which is not addressed here. In gamma spectroscopy, each emitted photon may remain unobserved or partially dectected (Compton effect); furthermore, some transitions can be non-radiative (e.g. occurring by electronic conversion). 
In our simplified scheme, the deexcitation process is purely radiative, and all the photons are fully detected: 
each event is then simply characterized by the list of transitions that occurred in the corresponding deexcitation cascade. For a given event, a specific gate is open if the corresponding transition is present in the list, 
and closed otherwise. A gate condition yields a set of selected events.

\begin{figure}
\includegraphics[scale=0.45]{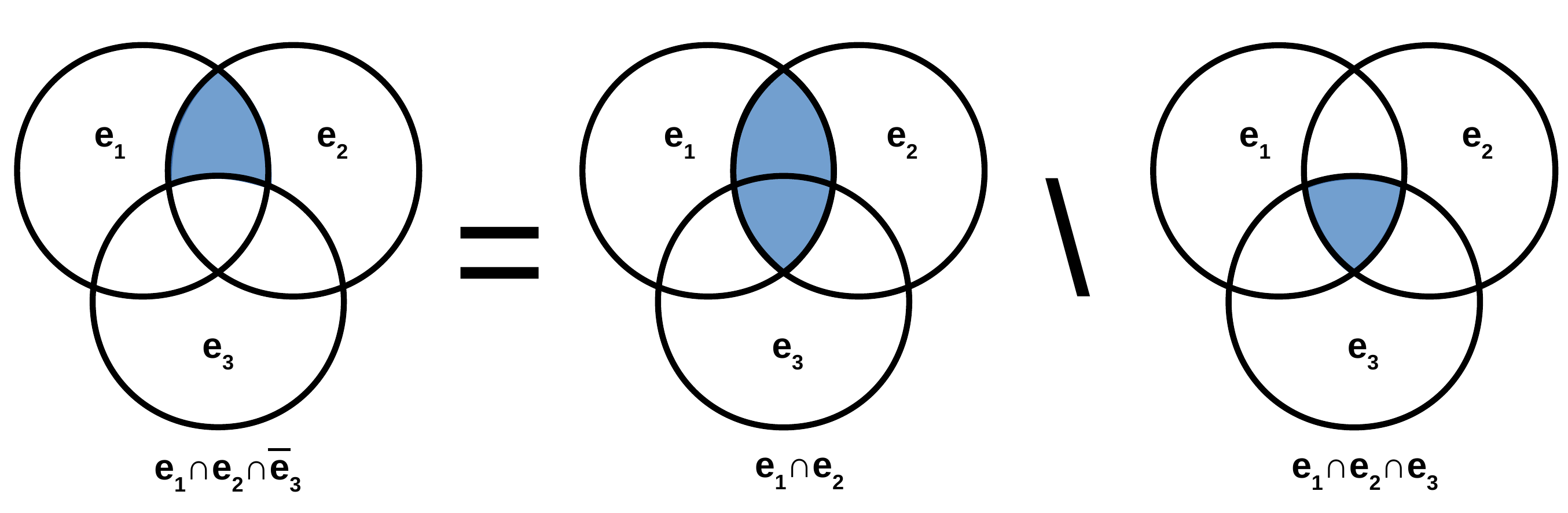}
\caption{Schematic illustration of the basic relation used to express exclusive elementary sets 
in terms of positive elementary sets.}
\label{Fig:set-difference}
\end{figure}

Let us first remind some conventional symbols and properties of set algebra:
\begin{itemize}
\item Set union: $E=E_1 \cup E_2$ 
contains all events that belong to $E_1$ and all events that belong to  $E_2$. 
The union of a series of sets reads:
\be 
E_1 \cup ... \cup E_n = \bigcup_{i=1}^n E_i \nn
\ee
\item Set intersection: $E=E_1\cap E_2$ 
contains all events that belong to both $E_1$ and $E_2$. 
The intersection of a series of sets reads:
\be
E_1\cap ... \cap E_n = \bigcap_{i=1}^n E_i \nn
\ee
\item Set difference: $E_1 \setminus E_2$
contains all events that belong to $E_1$ but not to $E_2$.
\item Set complement: $\bar{E}$
contains all events that do not belong to $E$;
it can be written $\bar{E} = U \setminus E$,
where $U$ is the universe of events 
(i.e. the set that contains all of them).
%; for instance, all the events that are recorded during an experiment). 
\item Intersection of a set $E_1$ with a set complement $\bar{E}_2$: 
\be 
E_1\cap \bar{E}_2 
&=& E_1 \cap [U \setminus E_2] 
= [E_1\cap U] \setminus [E_1\cap E_2] 
= E_1 \setminus [E_1\cap E_2] \nn
\ee
This relation will be particularly useful for the treatment of optional gate conditions.
It is illustrated by a graphical example in Figure~\ref{Fig:set-difference}.
\end{itemize}

Let us now present the different kinds of event sets we will deal with.
\begin{itemize}
\item A \textit{single set} $e$ is associated with single-gate conditions $G=\{g\}$.
\item A \textit{positive elementary set $E(G)$ of order N}
is associated with a positive explicit gate condition $G=\{g_1\cdot ...\cdot g_N\}$:
in short-hand notation, it is denoted by $E_{g_1...g_N}$.
It corresponds to the intersection of single sets $e_i=E(\{g_i\})$:
\be 
E_{g_1...g_N}&=& e_1\cap ... \cap e_N\nn
\ee
\item An \textit{exclusive elementary set $E(G)$ of order $(n,\bar{n})$} 
is associated with an exclusive explicit gate condition
$G=\{g^{(o)}_1\cdot ...\cdot g^{(o)}_n\cdot \bar{g}^{(c)}_1\cdot...\cdot\bar{g}^{(c)}_{\bar{n}}\}$:
in short-hand notation, it is denoted by 
$E_{g^{(o)}_1...g^{(o)}_n\bar{g}^{(c)}_1...\bar{g}^{(c)}_{\bar{n}}}$.
It corresponds to the intersection of single sets $e^{(o)}_i=E(\{g^{(o)}_i\})$
and single set complements $\bar{e}^{(c)}_i = \bar{E}(\{g^{(c)}_i\})$:
\be 
E_{g^{(o)}_1...g^{(o)}_n\bar{g}^{(c)}_1...\bar{g}^{(c)}_{\bar{n}}}
&=& e^{(o)}_1 \cap ... \cap e^{(o)}_n \cap \bar{e}^{(c)}_1 \cap ... \cap \bar{e}^{(c)}_{\bar{n}}\nn\\
&=& e^{(o)}_1 \cap ... \cap e^{(o)}_n 
\cap (U\setminus {e}^{(c)}_1) \cap ... \cap (U \setminus {e}^{(c)}_{\bar{n}})\nn
\ee
\item A \textit{combined set} $\mE(\mG)$ is associated with an optional gate condition
$\mG=\{g_1+...+g_N\}_m$.
It corresponds to the union of several elementary sets.
This case will be detailed later. 
\end{itemize}

\begin{figure}
\includegraphics[scale=0.45]{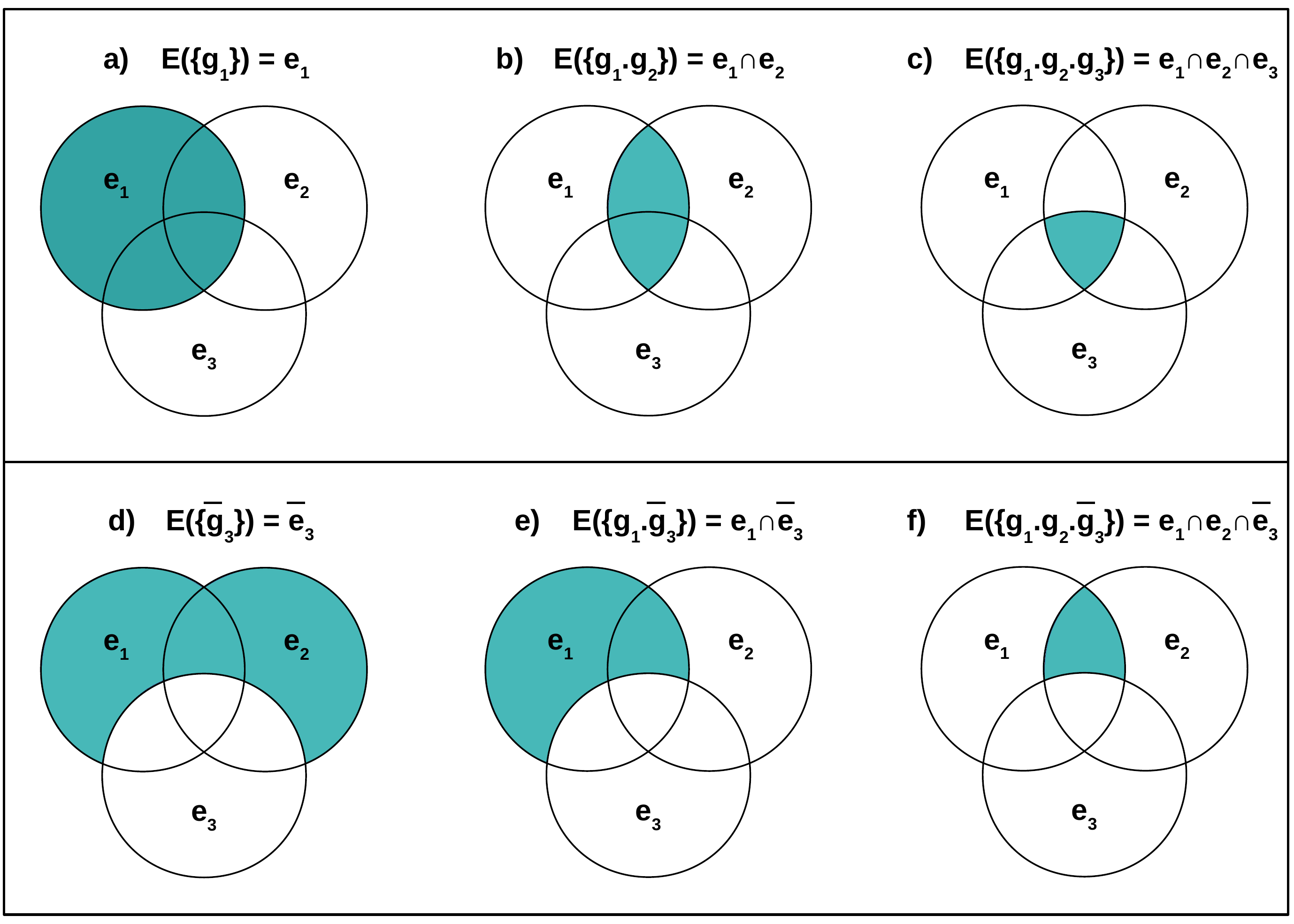}
\caption{Elementary sets: basic examples.
(a) Positive elementary set of order $1$ : $E(G=\{g_1\})$.
(b) Positive elementary set of order $2$ : $E(G=\{g_1\cdot g_2\})$.
(c) Positive elementary set of order $3$ : $E(G=\{g_1\cdot g_2\cdot g_3\})$.
(d) Exclusive elementary set of order $(0,1)$ : $E(G=\{\bar{g}_3\})$.
(e) Exclusive elementary set of order $(1,1)$ : $E(G=\{g_1 \cdot \bar{g}_3\})$.
(f) Exclusive elementary set of order $(2,1)$ : $E(G=\{g_1 \cdot g_2 \cdot \bar{g}_3\})$.
}
\label{Fig:elementary-set}
\end{figure}

\begin{figure}
\includegraphics[scale=0.3]{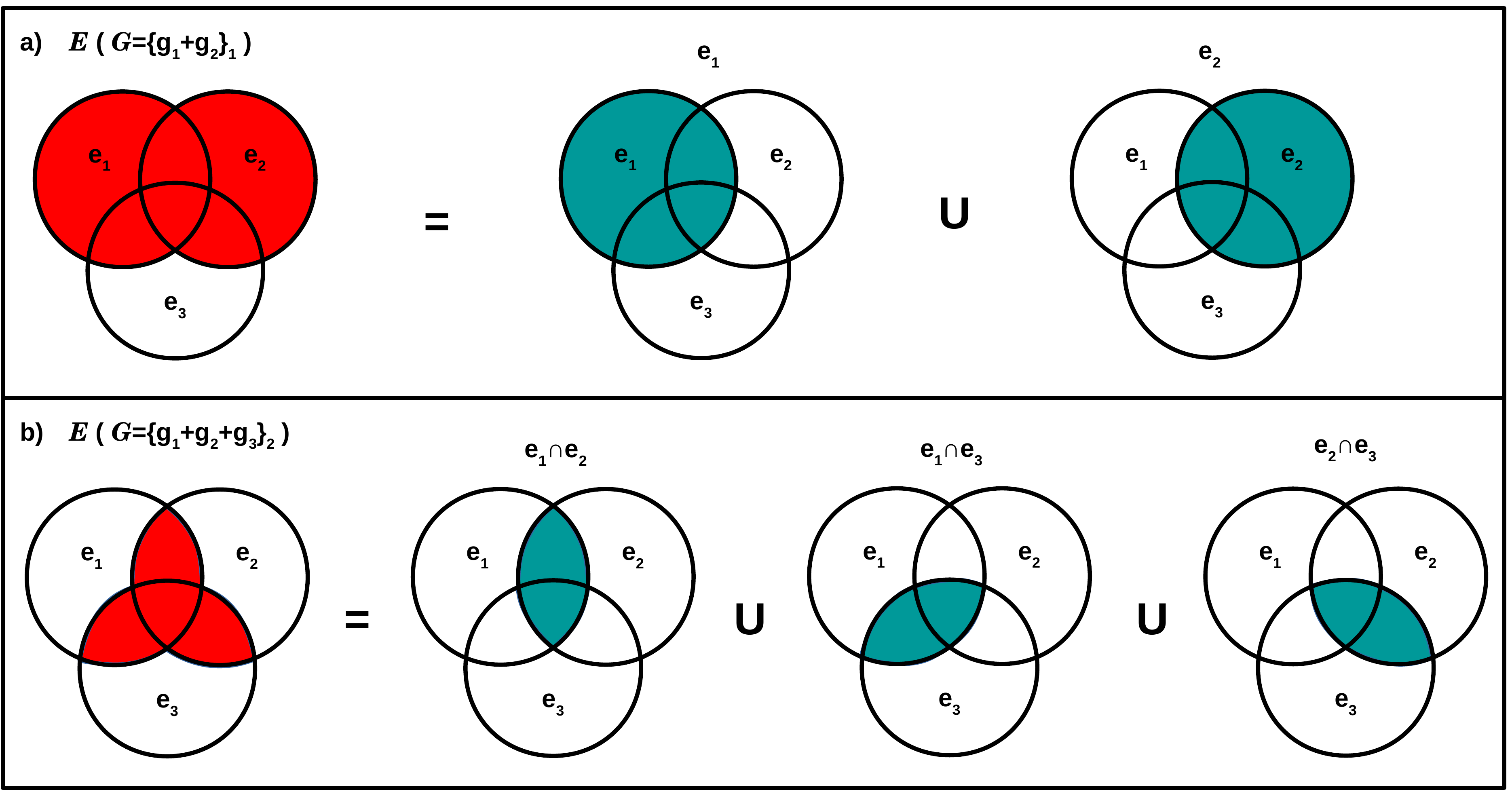}
\caption{Combined sets: basic examples.
(a) Combined set $\mE(\mG=\{g_1+g_2\}_1)$.
(b) Combined set $\mE(\mG=\{g_1+g_2+g_3\}_2)$.
}
\label{Fig:combined-set}
\end{figure}

Basic examples of elementary and combined sets 
are illustrated in Figures~\ref{Fig:elementary-set} and \ref{Fig:combined-set}.

\subsubsection{Associated spectra}

A spectrum is a histogram representation of events, 
displaying the number of photons counted in each energy interval. 
For each gate condition $G$, there is a \textit{set-spectrum} $S(E(G))$ 
representing the associated event set $E(G)$:
it gives the actual counting of photons emitted during the selected events.
For simplicity, $S(E(G))$ can be directly denoted by $S(G)$.

Note however that a spectrum does not necessarily provide a one-to-one representation of an event set:
other kinds of spectra can be obtained by combining set-spectra.
%Indeed, ordinary algebra operations can be applied on spectra, which can be multiplied by a number, summed and subtracted in the usual sense.
Let us consider for instance several set-spectra $S(E_i)$ representing event sets $E_i$.
A new spectrum $S$ can be obtained by performing a linear combination of $S(E_i)$
such as:
\be 
S&=&\sum_i c_i S(E_i) \nn
\ee
where the photon numbers of $S(E_i)$ are counted $c_i$ times (or subtracted $|c_i|$ times if $c_i<0$).
As a result, spectra can be constructed in such a way that some events are affected by multi-counting,
giving rise to the \textit{spiking effect} (artificial enhancement of some peaks).
We will call \textit{sum-spectrum} a spectrum of this kind.

For the present work, we need to define the following kinds of set-spectra:
\begin{itemize}
\item \textit{Positive elementary spectrum $\eS(G)$ of order $N$}, 
representing the event set $E(G)$ associated with a positive explicit condition $G=\{g_1\cdot ... \cdot g_N\}$.
In short-hand notation, it is denoted by $\eS_{g_1...g_N}$.
\item \textit{Exclusive elementary spectrum  $\eS(G)$ of order $(n,\bar{n})$}, 
representing the event set $E(G)$ associated with an exclusive explicit condition 
$G=\{g^{(o)}_1\cdot ... \cdot g^{(o)}_n \cdot \bar{g}^{(c)}_1\cdot ... \cdot \bar{g}^{(c)}_{\bar{n}}\}$.
In short-hand notation, it is denoted by:
$\eS_{g^{(o)}_1...g^{(o)}_n\bar{g}^{(c)}_1...\bar{g}^{(c)}_{\bar{n}}}$.
Building experimentally such a spectrum  would require combining gating and anti-gating methods 
as developed in Ref.~\cite{Stezowski99}.
%However, as detailed in Sec III.D.2, another approach, using the complementary between gates and anti-gates, has been followed in the present work.
However, in the present work, exclusive spectra are used as an intermediate step
and eventually expressed in terms of positive spectra, as detailed in Section~\ref{ssection-devel-positive-spectra}.
\item \textit{Combined spectrum} $\cS(\mG)$, representing a combined set $\mE(\mG)$  
associated with the optional condition $\mG=\{g_1+ ...+ g_N\}_m$.
We will see later how to express it as a combination of elementary spectra.
\end{itemize}
%In the rest of the discussion, if not otherwise mentionned, 
%we will call "spectrum" the representation of an event set (no multi-counting of events).

It is also useful to introduce a dedicated notation 
for specific kinds of sum-spectra, that will appear in later expressions: 
\begin{itemize}
\item \textit{Positive sum-spectrum of order $n$}, denoted by $\sigma(n,L)$, 
defined as the sum of spectra
associated with all positive explicit conditions of order $n$ 
that can be defined by picking $n$ gates in a given list $L=\{g_1,...,g_N\}$. 
It reads :
\be 
\sigma(n,L)&=&\sum_{\alpha=1}^{C^N_n} \eS(G_\alpha(n,L))\nn
\ee
\item \textit{Exclusive sum-spectrum of order $(n,\bar{n})$}, denoted by $\sigma(n,\bar{n},L)$, 
defined as the sum of spectra
associated with all exclusive explicit conditions of order  $(n,\bar{n})$
that can be defined by picking $n$ open gates and $\bar{n}$ closed gates in a given list $L=\{g_1,...,g_N\}$. 
It reads :
\be 
\sigma(n,\bar{n},L)&=&\sum_{\beta=1}^{C^N_{n+\bar{n}}\times C^{n+\bar{n}}_n} \eS(G_\beta(n,\bar{n},L))\nn
\ee
\item \textit{Spiked spectrum}: it is a usual kind of sum-spectrum, 
which is employed in practice as the simplest way to represent the events of a combined set.
Obtained by summing set-spectra of overlapping event sets, 
it involves some multi-counting, hence the "spiked" qualification.
The case of spiked spectra will be addressed in a dedicated subsection.
\end{itemize}

\subsubsection{Gamma-ray intensity and relative intensity}

Let us consider a given gamma ray emitted during the transition $\tVE_i$,
occurring with the probability $\PVE_i$
in the deexcitation cascade following nucleus formation:
\be 
\PVE_i&=&\frac{N_i}{N_{tot}}\nn,
\ee
where $N_{tot}$ is the total number of events 
(i.e. the number of nucleus formations followed by deexcitation)
and $N_i$ is the number of transitions $\tVE_i$ that occur.
Experimentally, a typical goal when a new transition $\tVE_i$ is observed 
is to quantify the probability $\PVE_i$ 
by measuring the corresponding peak size in a gamma-emission spectrum.
Usually, the studied spectra are subject to gate conditions
that make this peak more visible by reducing the background 
and the number of alternative cascades.
The purpose here is then to relate the peak size associated with $\tVE_i$ in a gated spectrum
to the emission probability $\PVE_i$.
Let us define the following quantities, 
for a given set of events $E$ associated with a gate condition $G$:
\begin{itemize}
\item \textit{Gated transition probability} $P_{\{G,i\}}$: 
probability for an event to verify condition $G$ and to contain transition $\tVE_i$. 
It corresponds to the ratio:
\be 
P_{\{G,i\}} &=& \frac{N_{\{G,i\}}}{N_{tot}}\nn
\ee
where $N_{\{G,i\}}$ is the number of events of $E(G)$ that involve $\tVE_i$.
This quantity will be expressed later as a function of the transition probability vector $\PV$ and matrix $\PM$.
\item \textit{Gated intensity} $I_i(G)$: 
fraction of counts in the gated spectrum that belong to the peak of $\tVE_i$.
It corresponds to the ratio:
\be 
I_i(G)&=&\frac{N_{\{G,i\}}}{N_{\{G,\gamma\}}}\nn
\ee
where $N_{\{G,\gamma\}}$ is the total number of gamma rays emitted during the events $E(G)$.
This number is directly given by the spectrum integral,
but it is not determined by the formalism we are developing here.
Indeed, $N_{\{G,\gamma\}}$ includes a number of emissions from the continuum of the level scheme,
which is not treated by our formalism (in its present version).
\item \textit{Relative gated intensity} $I^{(r)}_i(G)$: 
ratio between the peak sizes associated with $\tVE_i$
and with a reference transition $\tVE_{ref}$. 
It corresponds to:
\be 
I^{(r)}_i(G)&=&\frac{N_{\{G,i\}}}{N_{\{G,ref\}}} = \frac{P_{\{G,i\}}}{P_{\{G,ref\}}}\nn
\ee
where $N_{\{G,i\}}$ and $N_{\{G,ref\}}$ can be directly measured in the gated spectrum
%(under the above-mentionned simplifying assumptions that each transition emits a photon that is fully detected),
while $P_{\{G,i\}}$ and $P_{\{G,ref\}}$ can be expressed in terms of the transition probabilities
involved in vector $\PV$ and matrix $\PM$.
\end{itemize}
Let us finally define the \textit{relative intensity} $I^{(r)}_i$, 
which compares the occurrence of $\tVE_i$ and $\tVE_{ref}$ in the total set of events:
\be 
I^{(r)}_i&=&\frac{N_i}{N_{ref}}=\frac{\PVE_i}{\PVE_{ref}}\nn
\ee
This quantity is often given in the litterature 
to characterize the strength of a transition $\tVE_i$ observed in an experiment.
Let us note that it is in principle different from any gated relative intensity,
although the measurement of $I^{(r)}_i(G)$ is usually assumed to give an approximation of $I^{(r)}_i$.
Since the validity of such an approximation 
depends on the details of the gate condition and on the cascade structure, 
it is important to establish a quantitative relation between gated and ungated relative intensities,
which is the aim of this work.
%This will be realized in the present work, since both quantities can be expressed in terms of the discrete transition probabilities.

\subsection{Positive explicit gate condition ("and")}

As defined above, a positive explicit gate condition
consists in a list of gates that are all required to be open.
It is denoted by $G=\{g_1\cdot...\cdot g_N\}$,
and gives rise to a positive elementary spectrum $S(G)$
that represents the set of selected events $E(G)$.
We also specify that the gate list is ordered in such a way that 
the gates of lower indices correspond to transitions occurring earlier in the cascade, 
i.e. emitted by a higher energy level.
This will be symbolized by the relation: $g_1>...>g_N$.
Our purpose is now to express the gated probability $P_{\{G,i\}}$ of a transition $\tVE_i$ 
as a function of the transition probability vector $\PV$ and matrix $\PM$.
We remind that each element $\PVE_k$ of the transition probability vector 
gives the probability that transition $\tVE_k$ occurs during the deexcitation process
%of transition $\tVE_k$ to occur during the cascade, 
while each element $\PME_{ij}=\PME_{\tVE_i\ra \tVE_j}$ of the transition probability matrix $\PM$ 
gives the probability that, once transition $\tVE_i$ has occurred, 
it is followed by transition $\tVE_j$ after an arbitrary number of steps.
For the homogeneity of some expressions, we will also use the notation $P_{\tVE_k}=P_k$.
%The gated relative intensity being defined by $I^{(r)}_i(G)=P_{\{G,i\}}/P_{\{G,ref\}}$, 
%in the following we will focus on the expression of the gated transition probability $P_{\{G,i\}}$ 
%(probability that an event belongs to $E(G)$ and includes $\tVE_i$).

Let us start with examples for restricted numbers of gates $N$.
The shortest list is of course the \textbf{single gate}: $G=\{g_1\}$. 
A transition $\tVE_i$ that occurs in coincidence with $g_1$ 
can take place either "above" or "below" $g_1$ in the deexcitation cascade. 
Namely, "above $g_1$" means earlier in the cascade, and is denoted by $\tVE_i>g_1$;
"below $g_1$" means later in the cascade, and is denoted by $\tVE_i<g_1$. 
Depending on each case, the gated intensity $I_i(G)$ is expressed differently 
as a function of the transition probability vector $\PV$ and matrix $\PM$:
\begin{itemize}
\item If $\tVE_i>g_1$: $P_{\{G,i\}} = \PVE_{\tVE_i}\times \PME_{\tVE_i\ra g_1}$
\item If $g_1>\tVE_i$: $P_{\{G,i\}} = \PVE_{g_1}\times \PME_{g_1\ra \tVE_i}$
\end{itemize}
Globally, we can write:
\be 
P_{\{G,i\}}&=& P_{\tVE_i}\times \mP_{\tVE_i\ra g_1} + P_{g_1}\times \mP_{g_1\ra \tVE_i}\nn
\ee
since if $g_1>\tVE_i$ we have $\PME_{\tVE_i\ra g_1}=0$, 
and if $\tVE_i>g_1$ we have $\PME_{g_1\ra \tVE_i}=0$.

Let us now consider a \textbf{double gate} $G=\{g_1\cdot g_2\}$ (ordered as $g_1>g_2$):
\begin{itemize}
\item If $\tVE_i>g_1>g_2$: $P_{\{G,i\}}=\PVE_{\tVE_i}\times \PME_{\tVE_i\ra g_1}\times \PME_{g_1\ra g_2}$
\item If $g_1>\tVE_i>g_2$: $P_{\{G,i\}}=\PVE_{g_1}\times \PME_{g_1\ra \tVE_i}\times \PME_{\tVE_i\ra g_2}$
\item If $g_1>g_2>\tVE_i$: $P_{\{G,i\}}=\PVE_{g_1}\times \PME_{g_1\ra g_2}\times \PME_{g_2\ra \tVE_i}$
\end{itemize}
which corresponds to the global expression, where only one term is non-zero:
\be 
P_{\{G,i\}}&=& \PVE_{\tVE_i}\times \PME_{\tVE_i\ra g_1}\times \PME_{g_1\ra g_2}
+ \PVE_{g_1}\times \PME_{g_1\ra \tVE_i}\times \PME_{\tVE_i\ra g_2}
+\PVE_{g_1}\times \PME_{g_1\ra g_2}\times \PME_{g_2\ra \tVE_i}\nn
\ee

In order to generalize the expression of $P_{\{G,i\}}$, %to any positive explicit gate condition $G$,
let us introduce the \textit{transition cascade vectors} $\CV^h(G)$: 
$h$ indicates the position of $\tVE_i$ among the gates $g_x$.
In the following, the dependence of $\CV^h$ on $G$ will be implicit.
For $G=\{g_1\cdot g_2\}$, there are three possible cascade vectors:
\begin{itemize}
\item If $\tVE_i>g_1>g_2$: $\CV^0 = (\tVE_i,g_1,g_2)$
\item If $g_1>\tVE_i>g_2$: $\CV^1 = (g_1,\tVE_i,g_2)$
\item If $g_1>g_2>\tVE_i$: $\CV^2 = (g_1,g_2,\tVE_i)$
\end{itemize}
We will denote by $\CVE^h_k$ the transition associated with the component $k$ 
of the cascade vector $\CV^h$
(with the convention that $k$ starts from zero).
Now we can use $\CV^h$ to write $P_{\{G,i\}}$:
\be 
P_{\{G,i\}}&=&
\PVE_{\CVE^0_0}\times \PME_{\CVE^0_0\ra \CVE^0_1}\times \PME_{\CVE^0_1\ra \CVE^0_2}
+ \PVE_{\CVE^1_0}\times \PME_{\CVE^1_0\ra \CVE^1_1}\times \PME_{\CVE^1_1\ra \CVE^1_2}
+ \PVE_{\CVE^2_0}\times \PME_{\CVE^2_0\ra \CVE^2_1}\times \PME_{\CVE^2_1\ra \CVE^2_2}\nn\\
&=& \sum_{h=0}^2 \PVE_{\CVE^h_0} \times \prod_{j=1}^2 \PME_{\CVE^h_{j-1}\ra \CVE^h_{j}}\nn
\ee
%\be 
%P_{\{G,i\}}&=&
%\PVE_{\CVE^0_0}\times \PME_{\CVE^0_0\ra \CVE^0_1}\times \PME_{\CVE^0_1\ra \CVE^0_2}\nn\\
%&&+ \PVE_{\CVE^1_0}\times \PME_{\CVE^1_0\ra \CVE^1_1}\times \PME_{\CVE^1_1\ra \CVE^1_2}\nn\\
%&&+ \PVE_{\CVE^2_0}\times \PME_{\CVE^2_0\ra \CVE^2_1}\times \PME_{\CVE^2_1\ra \CVE^2_2}\nn\\
%&=& \sum_{h=0}^2 \PVE_{\CVE^h_0} \times \prod_{j=1}^2 \PME_{\CVE^h_{j-1}\ra \CVE^h_{j}}\nn
%\ee
This last expression can be easily generalized to a positive explicit gate condition
$G=\{g_1\cdot g_2 \cdot ... \cdot g_N\}$ implying \textbf{any number $N$ of gates}:
\be 
\label{Eq:eS-intensity}
P_{\{G,i\}}&=& \sum_{h=0}^N \PVE_{\CVE^h_0} \times \prod_{j=1}^N \PME_{\CVE^h_{j-1}\ra \CVE^h_j}
\ee

Note that, although the $h$ summation offers an elegant mathematical expression that is independent 
from the transition location in the cascade,
it will be more efficient in numerical calculation to determine for each considered transition $\tVE_i$
the corresponding position $h(\tVE_i)$ before performing the product (since all other $h$ terms are zero).

\subsection{Optional gate conditions ("or")}

With optional gate conditions, 
the list $L=\{g_1,...,g_N\}$ is a list of optional gates:
a minimal number $m$ of them is required to be open.
Such a condition is denoted by $\mG=\{g_1 +...+ g_N\}_m = L_{m/N}$, 
and is fulfilled every time a combination of at least $m$ gates among $L$ are open.
The set of events $\mE(\mG)$ that are selected by this condition
includes different elementary sets.
Indeed, for any list $L^{\prime}=\{g^{\prime}_1,...,g^{\prime}_n\}$ that is a sublist of $L$ with $n \geq m$,
the elementary set $E(\{g^{\prime}_1 \cdot...\cdot g^{\prime}_n\})$ is included in $\mE(\mG)$.
%all the elementary sets associated with an explicit list of open gates $L_{\prime}=\{g^{\prime}_1,...g^{\prime}_n\}$, as long as $L^\prime$ is a sublist of $L$ and $n \geq m$. 
Several such elementary sets have to be united in order to obtain $\mE(\mG)$,
hence the denomination of \textit{combined set}.
%, in contrast with the elementary sets associated with explicit conditions.

The most simple way to obtain $\mE(\mG)$ is to unite all the elementary sets 
corresponding to the minimal requirement of $m$ open gates.
Each such set is associated with a positive explicit gate condition of order $m$,
$G_\alpha(m,L)=\{g^{(\alpha)}_1\cdot ... \cdot g^{(\alpha)}_m \}$,
where $L^{(\alpha)} = \{g^{(\alpha)}_1, ... , g^{(\alpha)}_m \}$ is a sublist of $L$.
Note however that the spectrum $\cS(\mG)$ that represents the combined set $\mE(\mG)$
does not correspond to the sum of elementary spectra $\sum_\alpha S(G_\alpha)$. 
Indeed, the sets $E(G_\alpha)$ are overlapping, which means that one event of $\mE(\mG)$
can belong to several sets $E(G_\alpha)$,
%they will be counted several times in the spectrum summation,
%giving rise to the artificial enhancement of peak sizes called spiking effect.
giving rise to an artificial enhancement of peak sizes (spiking effect).

Let us illustrate this with a specific combined gate condition $\mG=\{g_1+g_2+g_3\}_2$.
Here, an event is selected if at least 2 gates are open, among a list of 3.
Namely, it has to fulfill at least one of the explicit gate conditions 
$\{g_1\cdot g_2\}$, or $\{g_1\cdot g_3\}$, or $\{g_2\cdot g_3\}$,
which define the elementary sets 
$E(\{g_1\cdot g_2\})=E_{g_1g_2}$, 
$E(\{g_1\cdot g_3\})=E_{g_1g_3}$ 
and $E(\{g_2\cdot g_3\})=E_{g_2g_3}$, respectively.
We can easily realize that these sets are overlapping:
indeed, every event for which the three optional gates are open
belongs to all elementary sets $E_{g_1g_2}$, $E_{g_1g_3}$ and $E_{g_2g_3}$.
As a result, such events are counted three times 
in the sum-spectrum $\eS_{g_1g_2} + \eS_{g_1g_3} + \eS_{g_2g_3}$.

Simple examples of optional conditions are detailed in Appendix \ref{App:spectrum-development-examples},
which can be consulted in parallel with the present subsection.

\subsubsection{Tiling of the combined set}

In order to avoid the spiking effect and obtain the combined spectrum $\cS(\mG)$ 
as an exact representation of the event set $\mE(\mG)$,
we have to express $\mE(\mG)$ as the union of elementary sets $E(G_\beta)$ that are not overlapping.
In other words, the elementary sets $E(G_\beta)$ that are considered
have to constitute a tiling of the combined set $\mE(\mG)$.
The non-overlapping criterion means that the gate conditions $G_\beta$ have to exclude each other:
this is possible only if the status of all the gates of $L$ 
is specified by each explicit condition $G_\beta$, 
using exclusive conditions in the cases where less than $N$ gates are required to be open.

We have seen that exclusive explicit conditions of order $(n,\bar{n})$ 
can be defined from the list $L=\{g_1,...g_N\}$; they are denoted by 
$G_\beta(n,\bar{n},L)
=\{g^{(\beta,o)}_1\cdot ... \cdot g^{(\beta,o)}_n 
\cdot \bar{g}^{(\beta,c)}_1\cdot ... \cdot \bar{g}^{(\beta,c)}_{\bar{n}}\}$,
where $L^{(\beta,o)}=\{g^{(\beta,o)}_1, ... g^{(\beta,o)}_n\}$ 
and $L^{(\beta,c)}=\{g^{(\beta,c)}_1, ... g^{(\beta,c)}_{\bar{n}}\}$
are sublists of $L$.
For the tiling of $\mE(\mG)$ by elementary sets, we need to consider all the conditions 
$G_\beta(n,\bar{n},L)$ such that $n\geq m$ and $n+\bar{n}=N$. 
For a given value of $n$, the number of $\beta$ combinations
is then given by $C^{N}_{N}\times C^{N}_n=C^N_n$.
This leads to the tiling relation :
\be 
\label{Eq:tiling}
\mE(\mG=L_{m/N})&=& \bigcup_{n=m}^N \bigcup_{\beta=1}^{C^N_n} E(G_\beta(n,N-n,L))
\ee
The tiling relation allows to obtain the combined spectrum representing $\mE(\mG)$ as:
\be 
\label{Eq:combined-spectrum}
\cS(\mG) &=& \sum_{n=m}^{N} \sum_{\beta=1}^{C^{N}_n} \eS(G_\beta(n,N-n,L))
= \sum_{n=m}^{N} \sigma(n,N-n,L)
\ee
where the second part of the equation
is obtained by replacing, for each value of $n$, the $\beta$ summation on elementary spectra
by the corresponding sum-spectrum $\sigma$.

Let us consider for instance the optional condition $\mG=\{g_1+g_2+g_3\}_2$. 
The tiling relation reads in this case:
\be 
\mE(\mG) &=& g_1g_2\bar{g}_3 \cup g_1g_3\bar{g}_2 \cup g_2g_3\bar{g}_1 \cup g_1g_2g_3 \nn
\ee
The combined spectrum is then given by:
\be 
\cS(\mG) &=& \eS_{g_1g_2\bar{g}_3} + \eS_{g_1g_3\bar{g}_2} + \eS_{g_2g_3\bar{g}_1} + \eS_{g_1g_2g_3}\nn
\ee
where each event of $\mE(\mG)$ is counted once and only once.
It involves the exclusive sum-spectra of order $(n,\bar{n})$ such that $n\geq 2$ and $n+\bar{n}=3$:
\be 
\sigma(2,1,\{g_1,g_2,g_3\})
&=& \eS_{g_1g_2\bar{g}_3} + \eS_{g_1g_3\bar{g}_2} + \eS_{g_2g_3\bar{g}_1} \nn\\
\sigma(3,0,\{g_1,g_2,g_3\}) 
&=& \sigma(3,\{g_1,g_2,g_3\}) = \eS_{g_1g_2g_3} \nn
\ee
in terms of which we can express the combined spectrum:
\be 
\cS(\mG=\{g_1+g_2+g_3\}_2) &=& \sigma(2,1,\{g_1,g_2,g_3\}) + \sigma(3,0,\{g_1,g_2,g_3\})\nn
\ee

\subsubsection{Development in positive elementary spectra}
\label{ssection-devel-positive-spectra}

We have seen that the tiling relation (\ref{Eq:tiling})
allows to obtain a combined spectrum $\cS(\mG)$ 
as a sum of elementary spectra $\eS(G_\beta)$.
However, this summation involves exclusive elementary spectra,
associated with conditions that impose gate closures.
In order to apply directly Eq.~(\ref{Eq:eS-intensity}) 
to establish the gamma-ray intensities in $\cS(\mG)$, 
we need to express the combined spectrum as a combination of positive elementary spectra.
From the experimentalist point of view, data analysis could involve anti-gating 
(namely, gates imposed to be closed);
however, one often prefers to construct spectra based on positive gating conditions.
Thus, the present approach will also allow a more classical correspondance 
between modelling and construction from an experimental data set.
%Actually, it has been shown e.g. in Ref.~\cite{Stezowski99} 
%that this is a promising alternative method to select for instance a specific rotational band.
%However, presently data analyses are usually based on positive gating conditions.
%The expression of the combined spectrum in terms of positive elementary spectra 
%will thus allow a more classical correspondance 
%between modelling and construction from an experimental data set.
%Furthermore, from the data analyst point of view, 
%one may prefer to avoid the construction of  spectra based on anti-gating conditions:

Let us first remind the expression of a set intersection with a complementary set:
$E=E_1\cap \bar{E}2=E_1\cap[U\setminus E_2]=E_1\setminus[E_1\cap E_2]$.
The corresponding spectrum is expressed as a subtraction:
\be 
S(E=E_1\cap \bar{E}2) &=& S(E_1)-S(E_1\cap E_2)\nn
\ee
This principle can be applied to spectra associated with exclusive explicit conditions.
The exclusive elementary set associated with a condition $G_\beta(n,\bar{n},L)$ is:
\be 
E_{g^{(\beta,o)}_1...g^{(\beta,o)}_n\bar{g}^{(\beta,c)}_1...\bar{g}^{(\beta,c)}_{\bar{n}}}
&=& e^{(\beta,o)}_1 \cap ... \cap e^{(\beta,o)}_n \cap \bar{e}^{(\beta,c)}_1 \cap ... 
\cap \bar{e}^{(\beta,c)}_{\bar{n}-1}\cap \bar{e}^{(\beta,c)}_{\bar{n}}\nn\\
&=& \left[e^{(\beta,o)}_1 \cap ... \cap e^{(\beta,o)}_n \cap \bar{e}^{(\beta,c)}_1 \cap ... \cap \bar{e}^{(\beta,c)}_{\bar{n}-1}\right]
\cap \left[U \setminus {e}^{(\beta,c)}_{\bar{n}} \right] \nn\\
&=&
\left[e^{(\beta,o)}_1 \cap ... \cap e^{(\beta,o)}_n \cap \bar{e}^{(\beta,c)}_1 \cap ... \cap \bar{e}^{(\beta,c)}_{\bar{n}-1}\right] \nn\\
&&\setminus
\left[
e^{(\beta,o)}_1 \cap ... \cap e^{(\beta,o)}_n \cap \bar{e}^{(\beta,c)}_1 \cap ... \cap \bar{e}^{(\beta,c)}_{\bar{n}-1} \cap {e}^{(\beta,c)}_{\bar{n}}
\right]\nn
\ee
so that the spectrum $S(G_\beta(n,\bar{n},L))$ can be decomposed as:
\be 
%\hspace*{-2cm}
S_{g^{(\beta,o)}_1...g^{(\beta,o)}_n\bar{g}^{(\beta,c)}_1...\bar{g}^{(\beta,c)}_{\bar{n}}}
&=& S_{g^{(\beta,o)}_1...g^{(\beta,o)}_n\bar{g}^{(\beta,c)}_1...\bar{g}^{(\beta,c)}_{\bar{n}-1}}
- S_{g^{(\beta,o)}_1...g^{(\beta,o)}_n\bar{g}^{(\beta,c)}_1...\bar{g}^{(\beta,c)}_{\bar{n}-1}g^{(\beta,c)}_{\bar{n}}}
\label{Eq:spectrum-develop-step1}
\ee
As a result, any exclusive elementary spectrum of order $(n,\bar{n})$
can be expressed as the combination of exclusive elementary spectra 
of order $(n,\bar{n}-1)$ and $(n+1,\bar{n}-1)$.
Applying this relation recursively,
we find that $S(G_\beta)$ can be developed as a combination of 
positive elementary spectra of order $p$, with $n\leq p \leq n+\bar{n}$.
To represent the process, it is convenient to use for an elementary spectrum 
the notation 
$S(G_\beta)=g^{(\beta,o)}_1...g^{(\beta,o)}_n\bar{g}^{(\beta,c)}_1...\bar{g}^{(\beta,c)}_{\bar{n}}$, 
and to symbolize the relation (\ref{Eq:spectrum-develop-step1}) by a factorisation:
\be 
S(G_\beta(n,\bar{n},L))&=&g^{(\beta,o)}_1...g^{(\beta,o)}_n\bar{g}^{(\beta,c)}_1...\bar{g}^{(\beta,c)}_{\bar{n}-1}(1-{g}^{(\beta,c)}_{\bar{n}})\nn
\ee
Along the recursive steps, each closed gate factor $\bar{g_x}$
is eventually replaced by the factor $(1-g_x)$,
leading to the following expression of $S(G_\beta)$ in terms of positive elementary spectra:
\be 
S(G_\beta(n,\bar{n},L))&=&g^{(\beta,o)}_1...g^{(\beta,o)}_n
(1-{g}^{(\beta,c)}_1)...(1-{g}^{(\beta,c)}_{\bar{n}})
\label{Eq:spectrum-develop-step2a}
\ee
Note that this expression makes use of the notation $g^{(\beta,c)}_i$ (with no bar on $g$)
indicating that the gate $i$ of the list $L^{(\beta,c)}$ has changed its original status from "closed" to "open".
Hence the distinction that was introduced in 
Section~\ref{sssection-def-gate-conditions}
for the definition of exclusive conditions: 
the lists $L^{(\beta,o)}$ and $L^{(\beta,c)}$ identify the gates involved in the condition, 
while the condition 
$G_\beta=\{g^{(\beta,o)}_1\cdot ... \cdot g^{(\beta,o)}_n
\cdot\bar{g}^{(\beta,c)}_1\cdot...\cdot\bar{g}^{(\beta,c)}_{\bar{n}}\}$
specifies their status, that vary when $S(G_\beta)$ is developed in terms of positive conditions.
Now, the development (\ref{Eq:spectrum-develop-step2a}) can be expressed as:
\be 
S(G_\beta(n,\bar{n},L))&=&
\sum_{p=n}^{n+\bar{n}} (-1)^{p-n} \sum_{\alpha=1}^{C^N_p} k_{\alpha,\beta} \; S(G_\alpha(p,L))
\label{Eq:spectrum-develop-step2}
\ee
where the $\alpha$ index identifies the different gate combinations $L^{(\alpha)}$
defining the positive explicit conditions $G_\alpha(p,L)$.
We have introduced the coefficient $k_{\alpha,\beta}$,
which is $1$ if the $\alpha$ combination appears in the development of $S(G_\beta)$, and $0$ otherwise.
The condition to have $k_{\alpha,\beta}=1$ is then:
\begin{itemize}
\item $L^{(\alpha)} \subseteq (L^{(\beta,o)}+L^{(\beta,c)})$ 
(all gates open in $G_\alpha$ are specified in $G_\beta$)
\item $L^{(\beta,o)} \subseteq L^{(\alpha)}$ 
(all gates open in $G_\beta$ are open in $G_\alpha$)
\end{itemize}
A given term $S(G_\alpha(p,L))$ appears in the development of several spectra $S(G_\beta(n,\bar{n},L))$.
Indeed, to determine a condition $G_\beta$ such that $k_{\alpha,\beta}\neq 0$:
\begin{itemize}
\item There are $C^p_n$ choices to pick the $n$ gates of $L^{(\alpha)}$ that belong to $L^{(\beta,o)}$
\item The remaining $p-n$ gates of $L^{(\alpha)}$ necessarily belong to $L^{(\beta,c)}$
\item The list $L^{(\beta,c)}$ contains $\bar{n}$ gates, of which $p-n$ belong to $L^{(\alpha)}$.
The remaining $\bar{n}-(n-p)$ closed gates of $G_\beta$ have to be picked from the $N-p$ gates of $L$
that are not specified by $G_{\alpha}$: 
there are $C^{N-p}_{\bar{n}-(p-n)}=C^{N-p}_{\bar{n}+n-p}$ possible combinations.
\end{itemize}
Thus, for fixed values of $n$, $\bar{n}$, $p$ and $\alpha$,
there are $C^p_n\times C^{N-p}_{\bar{n}+n-p}$ values of $\beta$ such that $k_{\alpha,\beta}=1$.
As a result, by performing a summation of Eq. (\ref{Eq:spectrum-develop-step2}) over the $\beta$ index
(for fixed values of $n$ and $\bar{n}$), we obtain:
\be 
\sum_\beta S(G_\beta(n,\bar{n},L))
&=& \sum_\beta \sum_{p=n}^{n+\bar{n}} (-1)^{p-n}\sum_\alpha k_{\alpha,\beta}\;S(G_\alpha(p,L))\nn\\
&=& \sum_{p=n}^{n+\bar{n}} (-1)^{p-n}\sum_\alpha S(G_\alpha(p,L)) \sum_\beta k_{\alpha,\beta}\nn\\
\sum_\beta S(G_\beta(n,\bar{n},L))
&=& \sum_{p=n}^{n+\bar{n}} (-1)^{p-n} (C^p_n\times C^{N-p}_{\bar{n}+n-p})\sum_\alpha S(G_\alpha(p,L))\nn
\ee
where the $\beta$ summation runs over the 
$C^N_{n+\bar{n}}\times C^{n+\bar{n}}_n$ possible combinations to define $G_{\beta}(n,\bar{n},L)$;
the $\alpha$ summation runs over the $C^N_p$ possible combinations to define $G_{\alpha}(p,L)$;
and for given values of $p$ and $\alpha$, we have
$\sum_\beta k_{\alpha,\beta}=C^p_n\times C^{N-p}_{\bar{n}+n-p}$.
For a given gate list $L=\{g_1,...,g_N\}$, 
the above equation gives the sum of all exclusive elementary spectra of order $(n,\bar{n})$ 
in terms of positive elementary spectra of order $p$ with $n\leq p \leq N$.
Replacing the $\beta$ and $\alpha$ summations by sum-spectra $\sigma$, 
this expression reads:
\be 
\sigma (n,\bar{n},L) &=& \sum_{p=n}^{n+\bar{n}} (-1)^{p-n} (C^p_n\times C^{N-p}_{\bar{n}+n-p})\; \sigma(p,L)
\label{Eq:sum-spectrum-relation}
\ee
Namely, for a given gate list $L=\{g_1,...,g_N\}$, 
the exclusive sum-spectrum of order $(n,\bar{n})$
can be expressed by a combination of positive sum-spectra of order $p$ 
with $n\leq p\leq n+\bar{n}$.

\begin{figure}
\includegraphics[scale=0.36]{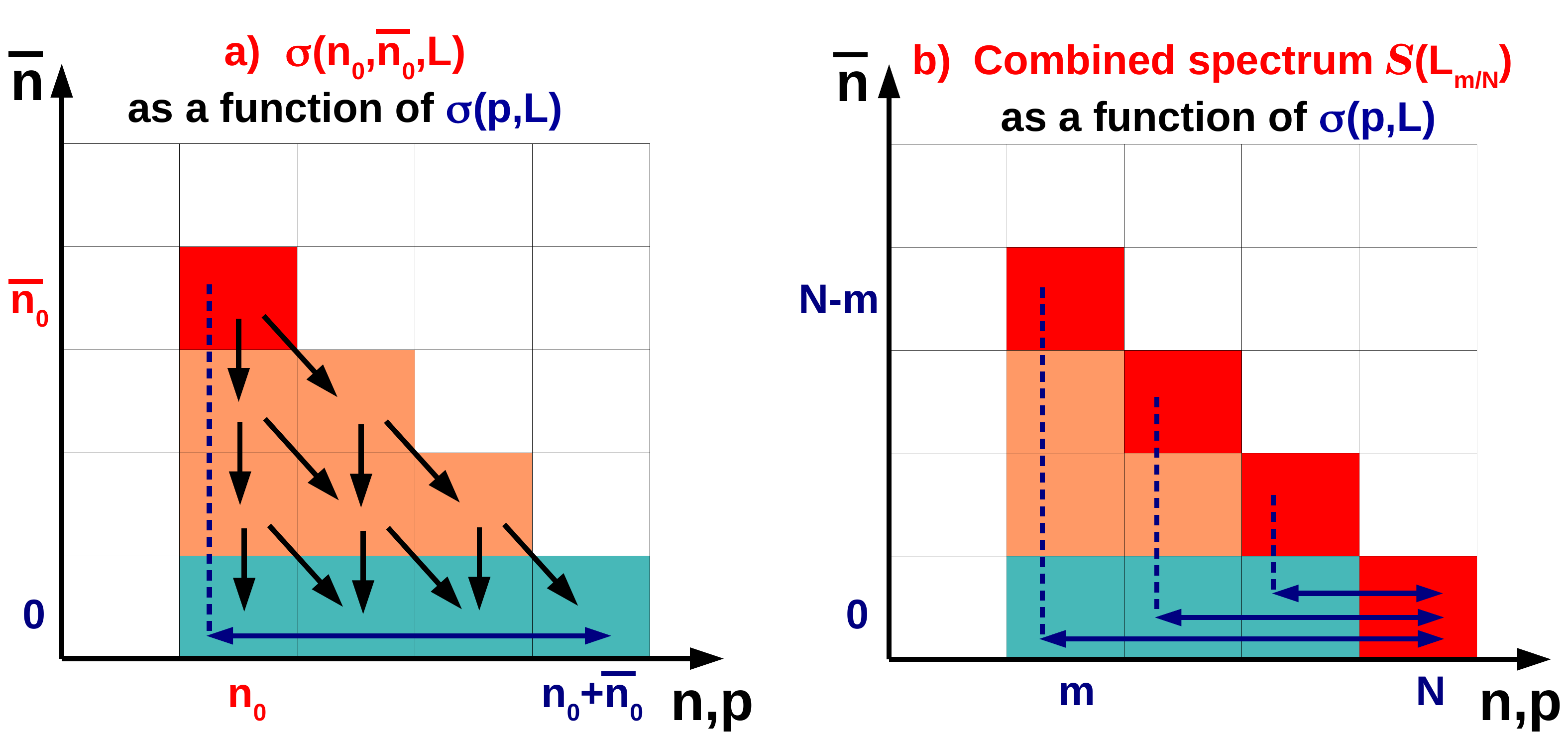}
\caption{(Color online) Schematic illustrations of the development in positive spectra. 
(a) Recursive development 
allowing to express an exclusive elementary spectrum of order $(n_0,\bar{n}_0)$
in terms of positive elementary spectra of order p, with $n_0\leq p\leq n_0+\bar{n}_0$.
The horizontal axis ($n$ or $p$) is the number of open gates;
the vertical axis ($\bar{n}$) is the number of closed gates.
One square of the grid represents an exclusive elementary spectrum of order $(n,\bar{n})$;
if $\bar{n}=0$, it is a positive elementary spectrum or order $p$.
The two arrows from each square $(n,\bar{n})$ 
symbolize its development in two elementary spectra 
of order $(n,\bar{n}-1)$ and $(n+1,\bar{n}-1)$, 
according to relation (\ref{Eq:spectrum-develop-step1}).
The horizontal arrow on the lower line 
shows the ensemble of positive elementary spectra of order $p$
involved in the development of the original elementary spectrum of order $(n_0,\bar{n}_0)$.
For correspondance with Eq.~(\ref{Eq:sum-spectrum-relation}),
we can also identify each square $(n,\bar{n})$ with the exclusive sum-spectrum $\sigma(n,\bar{n},L)$,
and each square $(p,0)$ with the positive sum-spectrum $\sigma(p,L)$.
(b) Similar illustration, applied to the combined spectrum $\cS(L_{m/N})$ 
composed of exclusive sum-spectra $\sigma(n,N-n,L)$ (red squares) 
according to the tiling relation $\cS(L_{m/N})=\sum_{n=m}^N\sigma(n,N-n,L)$
given by Eq.~(\ref{Eq:combined-spectrum}).
The arrows showing the steps of the recursive developments are not shown here.
Each horizontal arrow on the lower line illustrates
the expression of an exclusive sum-spectrum $\sigma(n,N-n,L)$ 
in terms of positive sum-spectra $\sigma(p,L)$,
corresponding to the development $\sigma(n,N-n,L)=\sum_{p=n}^N a_{n,p}\;\sigma(p,L)$ 
given by Eq.~(\ref{Eq:sum-spectrum-relation-2}).
Each square $\sigma(p,L)$ is involved in the development of all the squares 
$\sigma(n,N-n,L)$ with $n\leq p$,
hence the coefficients $c_m(p)=\sum_{n=m}^p a_{n,p}$ 
in the final development $\cS(L_{m/N})=\sum_{p=m}^N c_p(m)\;\sigma(p,L)$
given by Eq.~(\ref{Eq:combined-spectrum-development}).
}
\label{Fig:recurrence}
\end{figure}

This can be directly applied to the expression of the combined spectrum 
given by Eq.(\ref{Eq:combined-spectrum}), which results from the tiling relation.
Some simplifications occur due to the condition $n+\bar{n}=N$
(the exclusive explicit conditions involved in the tiling relation 
have to specify the status of every gate of $L$). 
We have in this case $C^{N-p}_{\bar{n}+n-p}=C^{N-p}_{N-p}=1$, 
%,and $C^N_{n+\bar{n}}=C^N_N=1$.
so the exclusive sum-spectra appearing in Eq.(\ref{Eq:combined-spectrum}) are given by:
\be 
\sigma (n,N-n,L) &=& \sum_{p=n}^{N} (-1)^{p-n} C^p_n \; \sigma(p,L)
= \sum_{p=n}^N a_{n,p}\;\sigma(p,L)
\label{Eq:sum-spectrum-relation-2}
\ee
where we have introduced the coefficients $a_{n,p}=(-1)^{p-n} C^p_n$.
The combined spectrum associated with 
the condition $\mG=L_{m/N}$ is expressed as the following combination:
\be 
\cS(L_{m/N})&=& \sum_{n=m}^N \sigma (n,N-n,L) \nn\\
&=& \sum_{n=m}^N \sum_{p=n}^{N} (-1)^{p-n} C^p_n \; \sigma(p,L) 
= \sum_{p=m}^N \sum_{n=m}^{p} (-1)^{p-n} C^p_n \; \sigma(p,L)
= \sum_{p=m}^N \sum_{n=m}^{p} a_{n,p} \; \sigma(p,L)\nn
\ee
%\be 
%\cS(L_{m/N})&=& \sum_{n=m}^N \sigma (n,N-n,L) \nn\\
%&=& \sum_{n=m}^N \sum_{p=n}^{N} (-1)^{p-n} C^p_n \; \sigma(p,L) \nn\\
%&=& \sum_{p=m}^N \sum_{n=m}^{p} (-1)^{p-n} C^p_n \; \sigma(p,L) \nn\\
%&=& \sum_{p=m}^N \sum_{n=m}^{p} a_{n,p} \; \sigma(p,L)\nn
%\ee
%where, for each $n$, the $\beta$ index runs over 
%$C^N_n$ combinations of $n$ open gates among the $N$ gates specified by $G_\beta$,
%and the $\alpha$ index runs over $C^N_p$ combinations of $p$ open gates among the list $L$.
The sum inversion is performed thanks to the relation verified by any function $f(n,p)$:
\be 
\sum_{
\begin{scriptsize}
\begin{array}{c}
n,p=m \\ 
n\leq p
\end{array} 
\end{scriptsize}
}^N f(n,p) &=& \sum_{n=m}^N\sum_{p=n}^N f(n,p)=\sum_{p=m}^N\sum_{n=m}^p f(n,p)\nn
\ee
Introducing the coefficients $c_p(m)=\sum_{n=m}^p a_{n,p}$ to express the linear combination,
the combined spectrum development in terms of positive elementary spectra reads:
\be 
%\cS(L_{m/N})&=& \sum_{p=m}^{N}\sum_{n=m}^p (-1)^{p-n} C^p_n \sum_\alpha S(G_\alpha(p,L)) \nn\\
\label{Eq:combined-spectrum-development}
\cS(L_{m/N})&=& \sum_{p=m}^N c_p(m)\; \sigma(p,L)
= \sum_{p=m}^N c_p(m) \sum_{\alpha=1}^{C^N_p} \eS(G_\alpha(p,L))
\ee
with
\be
c_p(m)&=& \sum_{n=m}^p a_{n,p} = \sum_{n=m}^p (-1)^{p-n} C^p_n \nn
%\\
%\label{Eq:coef_anp}
%a_{n,p}&=&(-1)^{p-n} C^p_n
\ee
The coefficients $c_p(m)$ involved in the development of $\cS(L_{m/N})$ 
can be easily obtained in practice by representing the coefficients $a_{n,p}=(-1)^{p-n}C^p_n$
in a universal table, where $n$ refers to the line number and $p$ to the column number.  
The binomial coefficients can even be recovered by hand, applying the Pascal relation
$C^{p+1}_{n+1}=C^{p}_n+C^{p}_{n+1}$ that allows to construst the well-known Pascal triangle.
The $a_{n,p}$ coefficients are shown in Table \ref{Tab:coefs} up to $p=10$.
%For the development of spectrum $\cS(L_{m/N})$,
%the coefficient $c_p(m)$ applied to the sum of positive elementary spectra of order $p$
The coefficients $c_p(m)$ are obtained by summing the elements of column $p$, starting at line $m$.
A schematic representation showing the principle of this development is shown in Figure~\ref{Fig:recurrence}.

\begin{table}
\begin{tabular}{|c|c|c|c|c|c|c|c|c|c|c}
\multicolumn{1}{c}{[p=1]} & \multicolumn{1}{c}{[p=2]} & \multicolumn{1}{c}{[p=3]} & \multicolumn{1}{c}{[p=4]} & \multicolumn{1}{c}{[p=5]} & \multicolumn{1}{c}{[p=6]} & \multicolumn{1}{c}{[p=7]} & \multicolumn{1}{c}{[p=8]} & \multicolumn{1}{c}{[p=9]} & \multicolumn{1}{c}{[p=10]} & \\ 
\cline{1-10} 
1   &   -2    &   3    &  -4    &   5   &   -6    &   7   &   -8   &    9   &  -10  & [n=1]	\\
\cline{1-10} 
\multicolumn{1}{c|}{} &	1   &   -3   &    6    &  -10   &  15  &   -21   &   28   &  -36   &   45  & [n=2]\\
\cline{2-10} 
\multicolumn{2}{c|}{} &	1    &  -4   &   10  &   -20   &   35   &  -56   &   84  &  -120 &  [n=3]\\
\cline{3-10} 
\multicolumn{3}{c|}{} &	1   &   -5    &  15    & -35    &  70   & -126   &  210 &  [n=4]\\
\cline{4-10} 
\multicolumn{4}{c|}{} &	 1   &   -6   &   21   &  -56   &  126  & -252  & [n=5]\\
\cline{5-10}
\multicolumn{5}{c|}{} &	1   &   -7    & 28   &  -84  &   210  & [n=6]\\
\cline{6-10}
\multicolumn{6}{c|}{} &	1   &   -8    &  36  &  -120 &  [n=7]\\
\cline{7-10}
\multicolumn{7}{c|}{} &	1    &  -9   &   45  & [n=8]\\
\cline{8-10}
\multicolumn{8}{c|}{} &	1  &   -10  & [n=9]\\
\cline{9-10}
\multicolumn{9}{c|}{} &	 1 &  [n=10]\\
\cline{10-10}
\end{tabular} 
\caption{
Coefficients $a_{n,p}$ that allow to express any combined spectrum $\cS(L_{m/N})$ with gate list size $N\leq 10$.}
\label{Tab:coefs}
\end{table}

\subsubsection{Gated intensity in a combined spectrum}

We consider here the combined spectrum $\cS(\mG)$
representing the set of events selected by the optional condition 
$\mG=L_{m/N}=\{g_1+...+g_N\}_m$.
Following Eq. (\ref{Eq:combined-spectrum-development}),
%obtained after application of the tiling relation (\ref{Eq:tiling}) 
%and development of the exclusive elementary spectra,
this combined spectrum is given by a linear combination of positive elementary spectra.
As a consequence, 
the number of events of $\mE(\mG)$ for which the transition $\tVE_i$ has occurred 
is given by a similar combination:
\be 
N_{\{\mG,i\}}&=& \sum_{p=m}^N c_p(m) \sum_{\alpha=1}^{C^N_p} N_{\{G_\alpha(p,L),i\}} \nn
\ee
as well as the gated probability of transition $\tVE_i$:
\be 
P_{\{\mG,i\}}&=& \frac{N_{\{\mG,i\}}}{N_{tot}} 
= \sum_{p=m}^N c_p(m) \sum_{\alpha=1}^{C^N_p} P_{\{G_\alpha(p,L),i\}} \nn
\ee
Each gated probability $P_{\{G_\alpha(p,L),i\}}$ 
associated with a positive explicit condition $G_\alpha(p,L)$
can be expressed in terms of the transition probabilities according to Eq.(\ref{Eq:eS-intensity}),
where the cascade vectors $\CV^h$ are determined by the list $L_\alpha$ 
of the corresponding gate combination.
We remind that the $h$ exponent gives the position of transition $\tVE_i$ 
among the sequence of gate transitions. 
For $G_\alpha=\{g^{(\alpha)}_1\cdot...\cdot g^{(\alpha)}_{p}\}$,
we have for instance $\CV^0(G_\alpha)=(\tVE_i,g^{(\alpha)}_1,...,g^{(\alpha)}_{p})$.
Note that, for fixed values of $p$ and $h$, 
the transition cascade vectors $\CV^h(G_\alpha)$ corresponding to the different $\alpha$ combinations
can be viewed as the different lines of a \textit{transition cascade matrix $\CM^h(p,L)$}.
%this matrix has $p+1$ columns (the elements of each vector $\CV^h(G_\alpha)$),
%and $C^N_p$ lines (the number of sublists of $p$ gates chosen in the list $L$).
An element $\CME^h_{\alpha,j}$ of this matrix
corresponds to the component $j$ of the cascade vector $T^h(G_\alpha)$.
The matrix $\CM^h(p,L)$ has $C^N_p$ lines and $p+1$ columns.
In the following, the dependence of each matrix $\CM^h$ on $p$ and $L$ will be implicit.
The gated probability $P_{\{\mG,i\}}$ is then given by:
\be 
\label{Eq:cS-intensity}
P_{\{\mG,i\}}&=& 
\sum_{p=m}^N c_p(m)
\left[{
\sum_{\alpha=1}^{C^N_p}
\left[{
\sum_{h=0}^p 
P_{\CME^h_{\alpha,0}} \times 
\prod_{j=1}^p \mP_{\CME^h_{\alpha,j-1}\ra \CME^h_{\alpha,j}}
}\right]
}\right]
\ee
%where, for each value of $p$, the $\alpha$ index runs over the $C^N_p$ combinations
%of $p$ gates taken from the list $L$ to define the condition positive explicit $G_\alpha(p,L)$.
The same formula applies to express the gated probability of a reference transition $\tVE_{ref}$,
so that the relative gated intensity $I^{(r)}_i(\mG)$ given by the ratio:
\be 
I^{(r)}_i(\mG)&=&\frac{N_{\{\mG,i\}}}{N_{\{\mG,ref\}}}=\frac{P_{\{\mG,i\}}}{P_{\{\mG,ref\}}}\nn
\ee
can be obtained either 
by measuring the peak areas $N_{\{\mG,i\}}$ and $N_{\{\mG,ref\}}$ in the combined spectrum $\cS(\mG)$,
or by implementing Eq. (\ref{Eq:cS-intensity}) 
to calculate the gated probabilities $P_{\{\mG,i\}}$ and $P_{\{\mG,ref\}}$.

\subsubsection{Case of "spiked" spectra}

The optional gate condition $\mG=\{g_1+...+g_N\}_m$ 
involves the list of gates $L=\{g_1,...,g_N\}$.
We have seen above that the simplest way to express the combined set $\mE(\mG)$
of events selected by $\mG$
is to perform the union of all positive elementary sets $E(G_\alpha)$,
where each condition $G_\alpha$ specifies a sublist of $m$ gates chosen among $L$:
\be 
\label{Eq:simple-union}
\mE(\mG)=\mE(L_{m/N}) &=& \bigcup_{\alpha=1}^{C^N_m} E(G_\alpha(m,L))
\ee
%where the $\alpha$ index runs over the $C^N_m$ possible combinations.
On the other hand, since the sets $E(G_\alpha)$ are overlapping,
multi-counting of events occurs if we want to represent $\mE(\mG)$
by the sum of elementary spectra $\eS(G_\alpha)$.
This is why the resulting spectrum is called a spiked spectrum:
\be 
\cSs(\mG)&=& \sum_{\alpha=1}^{C^N_m} \eS(G_\alpha(m,L)) \nn
\ee
%Using the sum-spectrum notation, this simply reads :
Note that the spiked spectrum is nothing but the sum-spectrum of order $m$:
\be 
\label{Eq:spiked-spectrum}
\cSs(\mG)&=&\sigma(m,L)
\ee

In the spiked spectrum, the (distorted) counting of the $\tVE_i$ transition is given by:
\be 
N_i(\cSs(\mG)) &=& \sum_{\alpha=1}^{C^N_m} N_{\{G_\alpha(m,L),i\}}
= \sum_{\alpha=1}^{C^N_m} N_{tot} \times P_{\{G_\alpha(m,L),i\}} \nn
\ee
where, as previously, the gated probabilities 
%associated with positive explicit conditions of order $m$
are given in terms of the transition probability vector $\PV$ and matrix $\PM$.
The involved elements are now indicated by the transition cascade matrices $\CM^h(m,L)$:
\be 
P_{\{G_\alpha(m,L),i\}}
&=& 
\sum_{h=0}^m 
P_{\CME^h_{\alpha,0}} \times 
\prod_{j=1}^m \mP_{\CME^h_{\alpha,j-1}\ra \CME^h_{\alpha,j}}
\nn
\ee
%\be 
%\sum_{\alpha=1}^{C^N_m} P_{\{G_\alpha(m,L),i\}}
%&=& \sum_{\alpha=1}^{C^N_m}
%\left[{
%\sum_{h=0}^N 
%P_{\CME^h_{\alpha,0}} \times 
%\prod_{j=1}^N \mP_{\CME^h_{\alpha,j-1}\ra \CME^h_{\alpha,j}}
%}\right] \nn
%\ee
The same formula applies to the reference transition $\tVE_{ref}$,
so that the \textit{spiked relative intensity} is obtained as:
\be 
I^{(r)}_i(\cSs(\mG))&=&\frac{N_i(\cSs(\mG))}{N_{ref}(\cSs(\mG))}
=\frac{\sum_\alpha P_{\{G_\alpha(m,L),i\}}}{\sum_\alpha P_{\{G_\alpha(m,L),ref\}}}\nn
\ee
We could conclude that,
although the spiked spectrum gives a distorted representation of the events selected by $\mG$,
it is also linked to the transition probability vector $\PV$ and matrix $\PM$ in a well-defined way.
So it can also be used as an analysis tool
if the goal is, for instance, to obtain information on the transition probabilities
by measuring the peak ratio ${N_i(\cSs(\mG))}/{N_{ref}(\cSs(\mG))}$.
Note however that, if one of the gates is a doubled transition
(namely, there is another possible transition with the same energy), 
even if we intend to apply an explicit condition,
the filtered events obey an effective condition that is combined: no elementary spectrum can be isolated.
In this case the analysis has to take into account the combinatory effects 
%that have been addressed in this section. 
associated woth "or"-type gate conditions.
The application of the present formalism to the case of doubled (or even multi-degenerate) transitions will be addressed in a future work.

\section{Application to a schematic level scheme}
\label{Sec:AppliSchematicLS}

In this section, we illustrate the predictive power of the analytic formula we have derived, 
both in the case of elementary and combined spectra. 
We have chosen to work on a schematic level scheme 
for the sake of clarity concerning the points we want to illustrate,
namely the difference between explicit and optional conditions, 
and the role of band communication.
This idealized approach allows to avoid obstacles such as the presence of degenerate transitions 
(which are not yet treated by the formalism). 
Furthermore, we can ignore the details of nucleus-formation mechanism and transition physical properties, 
thus ignoring constraints on the expected values of primary feeding and branching ratios
(this simplifies the choice of the input, but has no impact on the future applicability of the method to realistic cases).
We then remain with the problem of determining gated intensities for a list of transitions organized in a level scheme, 
with given emission probability and adjacency matrix. This will be done following the formalism presented in the previous sections.

\begin{figure}
\includegraphics[scale=0.6]{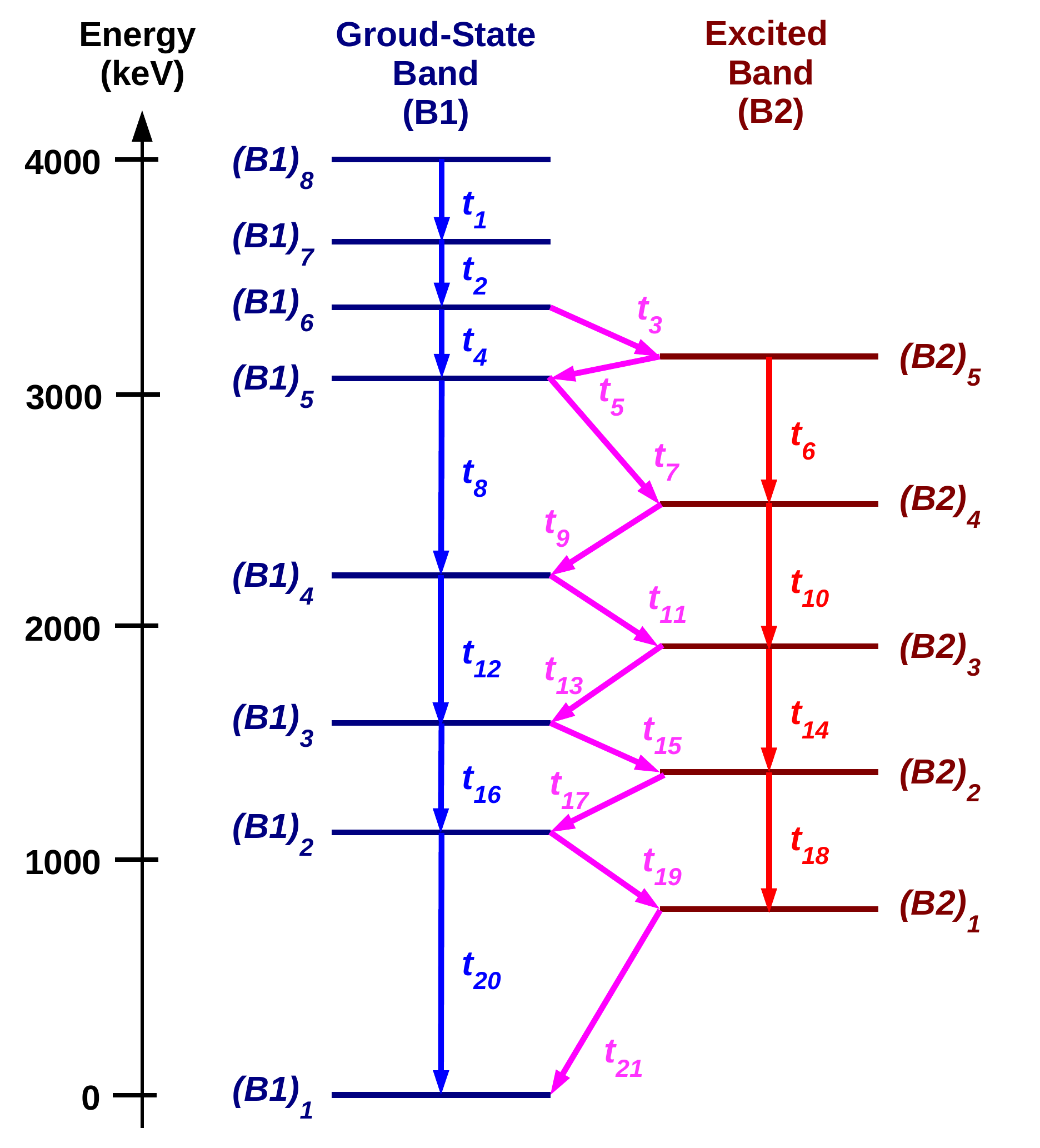}
\caption{(Color online) Schematic level scheme used to illustrate the predictive power of the formalism.
}
\label{Fig:scheme_ex13L}
\end{figure}

The schematic level scheme and corresponding transition scheme that are studied in this section 
are shown in Figure~\ref{Fig:scheme_ex13L}.
The level scheme is composed of two interacting structures, named "ground-state band" and "excited band". 
The transitions can link two successive levels in a given structure (intra-band transitions) 
or two neighboring levels of each structure (inter-band transitions).
This situation is quite usual in the structure of nuclei presenting for instance different kinds of deformation.
However, let us remind that no hypothesis is made here on the nature of the bands and transitions:
the role of electron conversion is ignored, and branching ratios are chosen arbitrarily.
Most importantly for our purpose, with this kind of level scheme, 
band communication allows different possibilities to pass from one transition to the other.
As a result, if we apply an optional gate condition using for instance the lower transitions of the ground-state band,
all the terms of the combination of sum-spectra given by Eq.~(\ref{Eq:combined-spectrum-development})
contribute to the gated intensity: this is what we need to check the full validity of this formula.

Let us remind the two possible approaches to describe the deexcitation process: characterization of the level space (list of levels with associated primary feeding and branching matrix), or characterization of the transition space (list of transitions with associated emission probabilities and adjacency matrix). 
As stated in Section~\ref{Sec:GraphTheory}, transition-space information can be deduced from level-space information.
For convenience, in our code, the original input concerns level-space information:
\begin{itemize}
\item List of levels, with corresponding level energy.
\item Branching matrix. In the case of a real application, this can be readily obtained from usual databases;
in our code, it was chosen by hand.
\item Primary feeding. 
For a given nucleus, this is not universal since it depends on the reaction mechanism: 
it should be determined for each studied experiment.
In our code, we have used an arbitrary function to distribute the primary feeding among the different levels.
\end{itemize}
The resulting level-space characterization is given by Table~\ref{Tab:ex13L_niveaux}
and the branching matrix presented in Table~\ref{Tab:ex13L_branching}.

\begin{table}
\begin{tabular}{|c|c|c|c|}
\hline 
Level Energy & Identification & Primary Feeding & Total Feeding \\
\hline 
\hline 
$0$ & $(B1)_1$ & $0.089$ & $1$\\
\hline 
$790$ & $(B2)_1$ & $0.048$ & $0.372$\\
\hline 
$1111$ & $(B1)_2$ & $0.097$ & $0.755$\\
\hline 
$1360$ & $(B2)_2$ & $0.049$ & $0.470$\\
\hline 
$1577$ & $(B1)_3$ & $0.099$ & $0.444$\\
\hline 
$1909$ & $(B2)_3$ & $0.050$ & $0.382$\\
\hline 
$2215$ & $(B1)_4$ & $0.099$ & $0.378$\\
\hline 
$2515$ & $(B2)_4$ & $0.049$ & $0.257$\\
\hline 
$3058$ & $(B1)_5$ & $0.096$ & $0.318$\\
\hline 
$3132$ & $(B2)_5$ & $0.048$ & $0.127$\\
\hline 
$3370$ & $(B1)_6$ & $0.094$ & $0.275$\\
\hline 
$3646$ & $(B1)_7$ & $0.092$ & $0.181$\\
\hline 
$4000$ & $(B1)_8$ & $0.089$ & $0.089$\\ 
\hline 
\end{tabular}
\caption{
Level-space characterization: list of levels with their denomination, energy, primary feeding and total feeding.
The total feeding is deduced from two kinds of input: 
primary feeding, and branching matrix given by Table~\ref{Tab:ex13L_branching}.
}
\label{Tab:ex13L_niveaux}
\end{table}

\begin{table}
%\hspace*{-3cm}
\begin{tabular}{|c|c|c|c|c|c|c|c|c|c|c|c|c|c|}
\hline
& $(B1)_1$ & $(B2)_1$ & $(B1)_2$ &  $(B2)_2$ & $(B1)_3$ & $(B2)_3$ & $(B1)_4$ & $(B2)_4$ & $(B1)_5$ & $(B2)_5$ & $(B1)_6$ & $(B1)_7$ & $(B1)_8$ \\
\hline 
$(B1)_1$ & $0$ & $0$ & $0$ & $0$ & $0$ & $0$ & $0$ & $0$ & $0$ & $0$ & $0$ & $0$ & $0$\\
\hline
$(B2)_1$ & $1$ & $0$ & $0$ & $0$ & $0$ & $0$ & $0$ & $0$ & $0$ & $0$ & $0$ & $0$ & $0$\\
\hline
$(B1)_2$ & $0.714$ & $0.286$ & $0$ & $0$ & $0$ & $0$ & $0$ & $0$ & $0$ & $0$ & $0$ & $0$ & $0$\\
\hline
$(B2)_2$ & $0$ & $0.231$ & $0.769$ & $0$ & $0$ & $0$ & $0$ & $0$ & $0$ & $0$ & $0$ & $0$ & $0$\\
\hline
$(B1)_3$ & $0$ & $0$ & $0.667$ & $0.333$ & $0$ & $0$ & $0$ & $0$ & $0$ & $0$ & $0$ & $0$ & $0$\\
\hline
$(B2)_3$ & $0$ & $0$ & $0$ & $0.714$ & $0.286$ & $0$ & $0$ & $0$ & $0$ & $0$ & $0$ & $0$ & $0$\\
\hline
$(B1)_4$ & $0$ & $0$ & $0$ & $0$ & $0.625$ & $0.375$ & $0$ & $0$ & $0$ & $0$ & $0$ & $0$ & $0$\\
\hline
$(B2)_4$ & $0$ & $0$ & $0$ & $0$ & $0$ & $0.741$ & $0.259$ & $0$ & $0$ & $0$ & $0$ & $0$ & $0$\\
\hline
$(B1)_5$ & $0$ & $0$ & $0$ & $0$ & $0$ & $0$ & $0.667$ & $0.333$ & $0$ & $0$ & $0$ & $0$ & $0$\\
\hline
$(B2)_5$ & $0$ & $0$ & $0$ & $0$ & $0$ & $0$ & $0$ & $0.800$ & $0.200$ & $0$ & $0$ & $0$ & $0$\\
\hline
$(B1)_6$ & $0$ & $0$ & $0$ & $0$ & $0$ & $0$ & $0$ & $0$ & $0.714$ & $0.286$ & $0$ & $0$ & $0$\\
\hline
$(B1)_7$ & $0$ & $0$ & $0$ & $0$ & $0$ & $0$ & $0$ & $0$ & $0$ & $0$ & $1$ & $0$ & $0$\\
\hline
$(B1)_8$ & $0$ & $0$ & $0$ & $0$ & $0$ & $0$ & $0$ & $0$ & $0$ & $0$ & $0$ & $1$ & $0$\\
\hline 
\end{tabular} 
\caption{
Level-space characterization: branching matrix $\BM$. 
Each element $\BME_{ij}$ gives the probability that level $i$ decays directly to level $j$.
}
\label{Tab:ex13L_branching}
\end{table}

The formalism that we use is based on a transition-space approach,
where the useful input is the transition probability vector $\PV$
and the adjacency matrix $\AM$.
Both can be deduced from the level-space input specified in Tables~\ref{Tab:ex13L_niveaux} and \ref{Tab:ex13L_branching}.
The elements of $\PV$ are given in Table~\ref{Tab:ex13L_transitions}, together with other transition properties;
the matrix $\AM$ is represented by Table~\ref{Tab:ex13L_AdjacencyMatrix}.
This is all the data needed to characterize the transition space and predict the profile of any kind of gated spectrum
obtained from the corresponding set of events. 

\begin{table}
\begin{tabular}{|c|c|c|c|c|}
\hline 
Identification & Emitting Level & Receiving Level & Transition Energy & Emission Probability \\
\hline
$\tVE_1$ & $(B1)_8$ & $(B1)_7$ & $354$ & $0.098$\\
\hline
$\tVE_2$ & $(B1)_7$ & $(B1)_6$ & $276$ & $0.199$\\
\hline
$\tVE_3$ & $(B1)_6$ & $(B2)_5$ & $238$ & $0.086$\\
\hline
$\tVE_4$ & $(B1)_6$ & $(B1)_5$ & $312$ & $0.216$\\
\hline
$\tVE_5$ & $(B2)_5$ & $(B1)_5$ & $74$ & $0.028$\\
\hline
$\tVE_6$ & $(B2)_5$ & $(B2)_4$ & $617$ & $0.111$\\
\hline
$\tVE_7$ & $(B1)_5$ & $(B2)_4$ & $543$ & $0.117$\\
\hline
$\tVE_8$ & $(B1)_5$ & $(B1)_4$ & $843$ & $0.233$\\
\hline
$\tVE_9$ & $(B2)_4$ & $(B1)_4$ & $300$ & $0.073$\\
\hline
$\tVE_{10}$ & $(B2)_4$ & $(B2)_3$ & $606$ & $0.209$\\
\hline
$\tVE_{11}$ & $(B1)_4$ & $(B2)_3$ & $306$ & $0.156$\\
\hline
$\tVE_{12}$ & $(B1)_4$ & $(B1)_3$ & $638$ & $0.259$\\
\hline
$\tVE_{13}$ & $(B2)_3$ & $(B1)_3$ & $332$ & $0.120$\\
\hline
$\tVE_{14}$ & $(B2)_3$ & $(B2)_2$ & $549$ & $0.299$\\
\hline
$\tVE_{15}$ & $(B1)_3$ & $(B2)_2$ & $217$ & $0.163$\\
\hline
$\tVE_{16}$ & $(B1)_3$ & $(B1)_2$ & $466$ & $0.325$\\
\hline
$\tVE_{17}$ & $(B2)_2$ & $(B1)_2$ & $249$ & $0.397$\\
\hline
$\tVE_{18}$ & $(B2)_2$ & $(B2)_1$ & $570$ & $0.119$\\
\hline
$\tVE_{19}$ & $(B1)_2$ & $(B2)_1$ & $321$ & $0.237$\\
\hline
$\tVE_{20}$ & $(B1)_2$ & $(B1)_1$ & $1111$ & $0.592$\\
\hline
$\tVE_{21}$ & $(B2)_1$ & $(B1)_1$ & $790$ & $0.408$\\
\hline 
\end{tabular} 
\caption{Transition-space characterization:
list of transitions with associated denomination, emitting level, receiving level, transition energy, and emission probability.
The data for emission probability can be considered either as an input, or as deduced from level-space information about primary feeding and branching ratios.
}
\label{Tab:ex13L_transitions}
\end{table}

\begin{table}
%\hspace*{-4cm}
%\begin{tiny}
%{\fontsize{5}{8}\selectfont 
\begin{tabular}{|c|c|c|c|c|c|c|c|c|c|c|c|c|c|c|c|c|c|c|c|c|c|}
\hline
& $\tVE_1$ & $\tVE_2$ & $\tVE_3$ & $\tVE_4$ & $\tVE_5$ & $\tVE_6$ & $\tVE_7$ & $\tVE_8$ & $\tVE_9$ & $\tVE_{10}$ & $\tVE_{11}$ & $\tVE_{12}$ & $\tVE_{13}$ & $\tVE_{14}$ & $\tVE_{15}$ & $\tVE_{16}$ & $\tVE_{17}$ & $\tVE_{18}$ & $\tVE_{19}$ & $\tVE_{20}$ & $\tVE_{21}$ \\
\hline
$\tVE_1$ &$0$ & $1$ & $0$ & $0$ & $0$ & $0$ & $0$ & $0$ & $0$ & $0$ & $0$ & $0$ & $0$ & $0$ & $0$ & $0$ & $0$ & $0$ & $0$ & $0$ & $0$\\
\hline
$\tVE_2$ &$0$ & $0$ & $0.286$ & $0.714$ & $0$ & $0$ & $0$ & $0$ & $0$ & $0$ & $0$ & $0$ & $0$ & $0$ & $0$ & $0$ & $0$ & $0$ & $0$ & $0$ & $0$\\
\hline
$\tVE_3$ &$0$ & $0$ & $0$ & $0$ & $0.200$ & $0.800$ & $0$ & $0$ & $0$ & $0$ & $0$ & $0$ & $0$ & $0$ & $0$ & $0$ & $0$ & $0$ & $0$ & $0$ & $0$\\
\hline
$\tVE_4$ &$0$ & $0$ & $0$ & $0$ & $0$ & $0$ & $0.333$ & $0.667$ & $0$ & $0$ & $0$ & $0$ & $0$ & $0$ & $0$ & $0$ & $0$ & $0$ & $0$ & $0$ & $0$\\
\hline
$\tVE_5$ &$0$ & $0$ & $0$ & $0$ & $0$ & $0$ & $0.333$ & $0.667$ & $0$ & $0$ & $0$ & $0$ & $0$ & $0$ & $0$ & $0$ & $0$ & $0$ & $0$ & $0$ & $0$\\
\hline
$\tVE_6$ &$0$ & $0$ & $0$ & $0$ & $0$ & $0$ & $0$ & $0$ & $0.259$ & $0.741$ & $0$ & $0$ & $0$ & $0$ & $0$ & $0$ & $0$ & $0$ & $0$ & $0$ & $0$\\
\hline
$\tVE_7$ &$0$ & $0$ & $0$ & $0$ & $0$ & $0$ & $0$ & $0$ & $0.259$ & $0.741$ & $0$ & $0$ & $0$ & $0$ & $0$ & $0$ & $0$ & $0$ & $0$ & $0$ & $0$\\
\hline
$\tVE_8$ &$0$ & $0$ & $0$ & $0$ & $0$ & $0$ & $0$ & $0$ & $0$ & $0$ & $0.375$ & $0.625$ & $0$ & $0$ & $0$ & $0$ & $0$ & $0$ & $0$ & $0$ & $0$\\
\hline
$\tVE_9$ &$0$ & $0$ & $0$ & $0$ & $0$ & $0$ & $0$ & $0$ & $0$ & $0$ & $0.375$ & $0.625$ & $0$ & $0$ & $0$ & $0$ & $0$ & $0$ & $0$ & $0$ & $0$\\
\hline
$\tVE_{10}$ &$0$ & $0$ & $0$ & $0$ & $0$ & $0$ & $0$ & $0$ & $0$ & $0$ & $0$ & $0$ & $0.286$ & $0.714$ & $0$ & $0$ & $0$ & $0$ & $0$ & $0$ & $0$\\
\hline
$\tVE_{11}$ &$0$ & $0$ & $0$ & $0$ & $0$ & $0$ & $0$ & $0$ & $0$ & $0$ & $0$ & $0$ & $0.286$ & $0.714$ & $0$ & $0$ & $0$ & $0$ & $0$ & $0$ & $0$\\
\hline
$\tVE_{12}$ &$0$ & $0$ & $0$ & $0$ & $0$ & $0$ & $0$ & $0$ & $0$ & $0$ & $0$ & $0$ & $0$ & $0$ & $0.333$ & $0.667$ & $0$ & $0$ & $0$ & $0$ & $0$\\
\hline
$\tVE_{13}$ &$0$ & $0$ & $0$ & $0$ & $0$ & $0$ & $0$ & $0$ & $0$ & $0$ & $0$ & $0$ & $0$ & $0$ & $0.333$ & $0.667$ & $0$ & $0$ & $0$ & $0$ & $0$\\
\hline
$\tVE_{14}$ &$0$ & $0$ & $0$ & $0$ & $0$ & $0$ & $0$ & $0$ & $0$ & $0$ & $0$ & $0$ & $0$ & $0$ & $0$ & $0$ & $0.769$ & $0.231$ & $0$ & $0$ & $0$\\
\hline
$\tVE_{15}$ &$0$ & $0$ & $0$ & $0$ & $0$ & $0$ & $0$ & $0$ & $0$ & $0$ & $0$ & $0$ & $0$ & $0$ & $0$ & $0$ & $0.769$ & $0.231$ & $0$ & $0$ & $0$\\
\hline
$\tVE_{16}$ &$0$ & $0$ & $0$ & $0$ & $0$ & $0$ & $0$ & $0$ & $0$ & $0$ & $0$ & $0$ & $0$ & $0$ & $0$ & $0$ & $0$ & $0$ & $0.286$ & $0.714$ & $0$\\
\hline
$\tVE_{17}$ &$0$ & $0$ & $0$ & $0$ & $0$ & $0$ & $0$ & $0$ & $0$ & $0$ & $0$ & $0$ & $0$ & $0$ & $0$ & $0$ & $0$ & $0$ & $0.286$ & $0.714$ & $0$\\
\hline
$\tVE_{18}$ &$0$ & $0$ & $0$ & $0$ & $0$ & $0$ & $0$ & $0$ & $0$ & $0$ & $0$ & $0$ & $0$ & $0$ & $0$ & $0$ & $0$ & $0$ & $0$ & $0$ & $1$\\
\hline
$\tVE_{19}$ &$0$ & $0$ & $0$ & $0$ & $0$ & $0$ & $0$ & $0$ & $0$ & $0$ & $0$ & $0$ & $0$ & $0$ & $0$ & $0$ & $0$ & $0$ & $0$ & $0$ & $1$\\
\hline
$\tVE_{20}$ &$0$ & $0$ & $0$ & $0$ & $0$ & $0$ & $0$ & $0$ & $0$ & $0$ & $0$ & $0$ & $0$ & $0$ & $0$ & $0$ & $0$ & $0$ & $0$ & $0$ & $0$\\
\hline
$\tVE_{21}$ &$0$ & $0$ & $0$ & $0$ & $0$ & $0$ & $0$ & $0$ & $0$ & $0$ & $0$ & $0$ & $0$ & $0$ & $0$ & $0$ & $0$ & $0$ & $0$ & $0$ & $0$\\
\hline 
\end{tabular} 
%\end{tiny}
%}
\caption{Transition-space characterization: adjacency matrix $\AM$. 
Each element $\AME_{ij}$ gives the probability that transition $i$ is immediately followed by transition $j$.
These numbers can be considered either as an input, or as deduced from level-space information about branching ratios.
}
\label{Tab:ex13L_AdjacencyMatrix}
\end{table}

\subsection{Comparison of analytical and numerical gated spectra}

\begin{figure}
\includegraphics[scale=0.65]{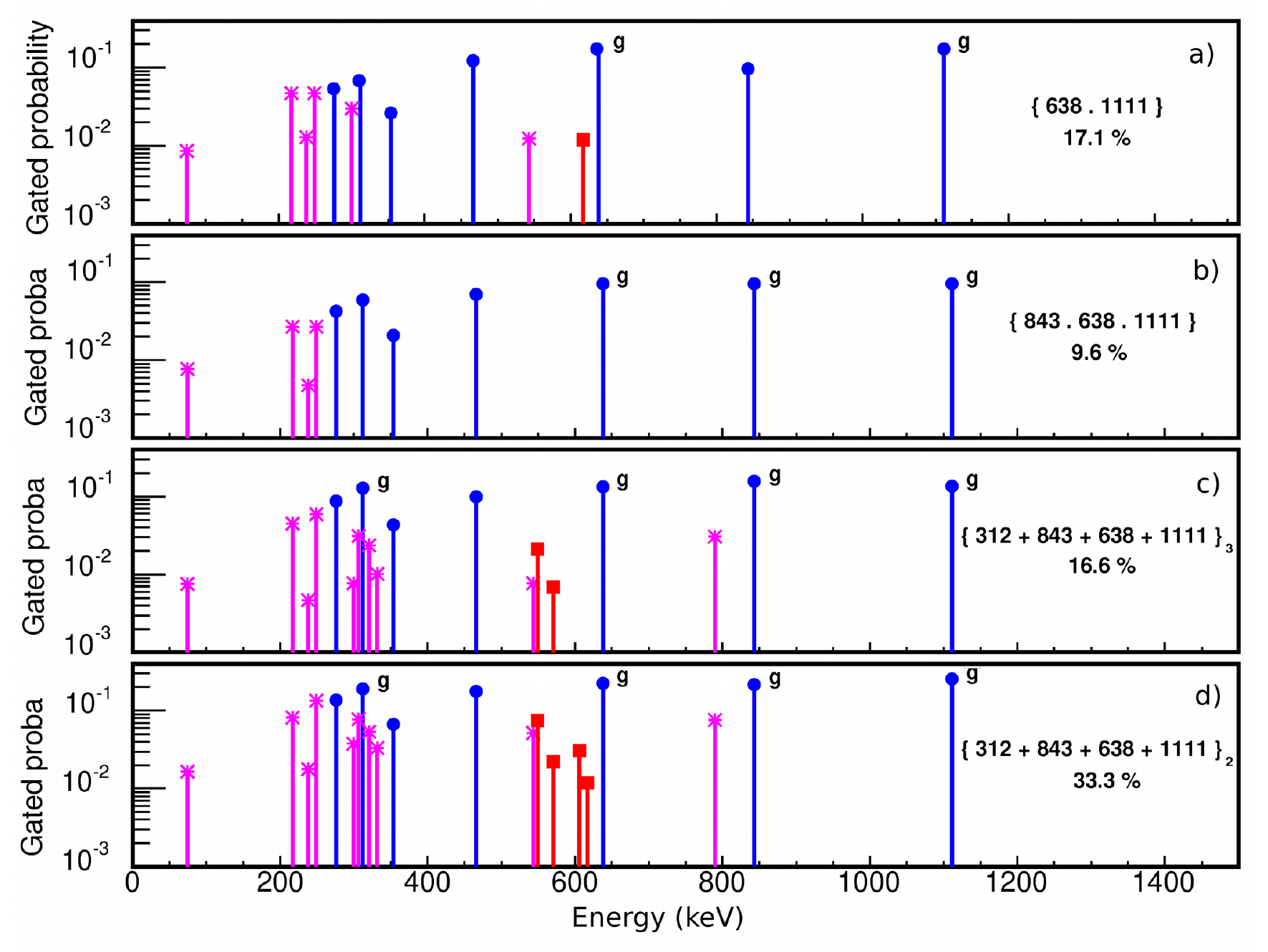}
\caption{(Color online) Gated spectra associated with the schematic level scheme of Figure~\ref{Fig:scheme_ex13L},
with transition-space properties given by Tables \ref{Tab:ex13L_transitions} and \ref{Tab:ex13L_AdjacencyMatrix}.
Gate conditions are specified on each panel : 
positive explicit conditions 
$G_1=\{\tVE_{12}\cdot \tVE_{20}\}$ for panel a), 
$G_2=\{\tVE_{8}\cdot \tVE_{12}\cdot \tVE_{20}\}$ for panel b); 
optional conditions 
$G_3=\{\tVE_{4} + \tVE_{8} + \tVE_{12} + \tVE_{20}\}_3$ for panel c), 
$G_4=\{\tVE_{4} + \tVE_{8} + \tVE_{12} + \tVE_{20}\}_2$ for panel d).
The lines correspond to numerical results, 
and the markers to analytical results 
(dots: intra-band transitions of the ground-state band;
squares: intra-band transitions of the excited band;
stars: inter-band transitions).
The vertical axis corresponds to the gated probability of each transition.
The percentage indicated on each panel is the probability that an event belongs to the selected set.
Transitions used as gates are indicated by the letter "g".
}
\label{Fig:spectra_ex13L}
\end{figure}

We now operate the presented formalism to obtain different gated spectra. 
The first step is to obtain the probability matrix $\PM$
by applying Eq.~(\ref{Eq:MatriceProba}):
the result is presented in Table~\ref{Tab:ex13L_ProbabilityMatrix}.
Next, we have to specify a gate condition $G$
and run an algorithm that yields the gated probability $P_{\{G,i\}}$ of each transition $\tVE_i$.
In the case of a positive explicit gate condition $G=\{g_1 \cdot ... \cdot g_N\}$,
these numbers are given by a straightforward application of Eq.~(\ref{Eq:eS-intensity}).
Note that transitions used as gates also have a gated probability attributed:
it corresponds to the probability that an event belongs to the selected set,
given by
\be 
P_{\{G\}} &=& \PVE_{g_1} \times \prod_{j=2}^{N} \PME_{g_{j-1}\ra g_{j}}\nn
\ee
We then obtain an elementary spectrum such as 
those represented on the two upper panels of Figure~\ref{Fig:spectra_ex13L}.
In the case of an optional gate condition $\mG=\{g_1+...+g_N\}_m=L_{m/N}$, 
we have to apply Eq.~(\ref{Eq:cS-intensity}), which requires several steps.
Starting from an empty combined spectrum,
for each given value of $p$ such that $m\leq p \leq N$, we have to: 
\begin{enumerate}
\item Determine the $C^N_p$ combinations of gates that will define the positive explicit conditions $G_\alpha(p,L)$
(where $\alpha$ identifies each combination). In practice, we calculate a combination matrix where each line $\alpha$
gives a sub-list of $p$ gates, identified by their position in the list $L$.
\item Sum the $C^N_p$ elementary spectra obtained by application of Eq.~(\ref{Eq:eS-intensity})
with the sets of gates $G_\alpha(p,L)$ given by each line of the combination matrix.
This gives the sum-spectrum $\sigma(p,L)$.
\item Calculate the coefficient $c_p(m)=\sum_{n=m}^p (-1)^{(p-n)}C^p_n$.
\item Add to the combined spectrum the sum-spectrum $\sigma(p,L)$ affected by the factor $c_p(m)$.
\end{enumerate}
We then obtain a combined spectrum such as 
those represented on the two lower panels of Figure~\ref{Fig:spectra_ex13L}.

\begin{table}
%\hspace*{-4.5cm}
%\begin{tiny}
\begin{tabular}{|c|c|c|c|c|c|c|c|c|c|c|c|c|c|c|c|c|c|c|c|c|c|}
\hline
& $\tVE_1$ & $\tVE_2$ & $\tVE_3$ & $\tVE_4$ & $\tVE_5$ & $\tVE_6$ & $\tVE_7$ & $\tVE_8$ & $\tVE_9$ & $\tVE_{10}$ & $\tVE_{11}$ & $\tVE_{12}$ & $\tVE_{13}$ & $\tVE_{14}$ & $\tVE_{15}$ & $\tVE_{16}$ & $\tVE_{17}$ & $\tVE_{18}$ & $\tVE_{19}$ & $\tVE_{20}$ & $\tVE_{21}$ \\
\hline 
$\tVE_1$ & $0$ & $1$ & $0.286$ & $0.714$ & $0.057$ & $0.229$ & $0.257$ & $0.514$ & $0.126$ & $0.360$ & $0.240$ & $0.400$ & $0.171$ & $0.428$ & $0.191$ & $0.381$ & $0.476$ & $0.143$ & $0.245$ & $0.612$ & $0.388$\\
\hline
$\tVE_2$ & $0$ & $0$ & $0.286$ & $0.714$ & $0.057$ & $0.229$ & $0.257$ & $0.514$ & $0.126$ & $0.360$ & $0.240$ & $0.400$ & $0.171$ & $0.428$ & $0.191$ & $0.381$ & $0.476$ & $0.143$ & $0.245$ & $0.612$ & $0.388$\\
\hline
$\tVE_3$ & $0$ & $0$ & $0$ & $0$ & $0.200$ & $0.800$ & $0.067$ & $0.133$ & $0.225$ & $0.642$ & $0.134$ & $0.224$ & $0.222$ & $0.554$ & $0.149$ & $0.297$ & $0.541$ & $0.162$ & $0.239$ & $0.598$ & $0.402$\\
\hline
$\tVE_4$ & $0$ & $0$ & $0$ & $0$ & $0$ & $0$ & $0.333$ & $0.667$ & $0.086$ & $0.247$ & $0.282$ & $0.471$ & $0.151$ & $0.378$ & $0.207$ & $0.415$ & $0.450$ & $0.135$ & $0.247$ & $0.618$ & $0.382$\\
\hline
$\tVE_5$ & $0$ & $0$ & $0$ & $0$ & $0$ & $0$ & $0.333$ & $0.667$ & $0.086$ & $0.247$ & $0.282$ & $0.471$ & $0.151$ & $0.378$ & $0.207$ & $0.415$ & $0.450$ & $0.135$ & $0.247$ & $0.618$ & $0.382$\\
\hline
$\tVE_6$ & $0$ & $0$ & $0$ & $0$ & $0$ & $0$ & $0$ & $0$ & $0.259$ & $0.741$ & $0.097$ & $0.162$ & $0.239$ & $0.599$ & $0.134$ & $0.268$ & $0.563$ & $0.169$ & $0.237$ & $0.594$ & $0.406$\\
\hline
$\tVE_7$ & $0$ & $0$ & $0$ & $0$ & $0$ & $0$ & $0$ & $0$ & $0.259$ & $0.741$ & $0.097$ & $0.162$ & $0.239$ & $0.599$ & $0.134$ & $0.268$ & $0.563$ & $0.169$ & $0.237$ & $0.594$ & $0.406$\\
\hline
$\tVE_8$ & $0$ & $0$ & $0$ & $0$ & $0$ & $0$ & $0$ & $0$ & $0$ & $0$ & $0.375$ & $0.625$ & $0.107$ & $0.268$ & $0.244$ & $0.488$ & $0.394$ & $0.118$ & $0.252$ & $0.630$ & $0.370$\\
\hline
$\tVE_9$ & $0$ & $0$ & $0$ & $0$ & $0$ & $0$ & $0$ & $0$ & $0$ & $0$ & $0.375$ & $0.625$ & $0.107$ & $0.268$ & $0.244$ & $0.488$ & $0.394$ & $0.118$ & $0.252$ & $0.630$ & $0.370$\\
\hline
$\tVE_{10}$ & $0$ & $0$ & $0$ & $0$ & $0$ & $0$ & $0$ & $0$ & $0$ & $0$ & $0$ & $0$ & $0.286$ & $0.714$ & $0.095$ & $0.190$ & $0.623$ & $0.187$ & $0.232$ & $0.581$ & $0.419$\\
\hline
$\tVE_{11}$ & $0$ & $0$ & $0$ & $0$ & $0$ & $0$ & $0$ & $0$ & $0$ & $0$ & $0$ & $0$ & $0.286$ & $0.714$ & $0.095$ & $0.190$ & $0.623$ & $0.187$ & $0.232$ & $0.581$ & $0.419$\\
\hline
$\tVE_{12}$ & $0$ & $0$ & $0$ & $0$ & $0$ & $0$ & $0$ & $0$ & $0$ & $0$ & $0$ & $0$ & $0$ & $0$ & $0.333$ & $0.667$ & $0.256$ & $0.077$ & $0.264$ & $0.659$ & $0.341$\\
\hline
$\tVE_{13}$ & $0$ & $0$ & $0$ & $0$ & $0$ & $0$ & $0$ & $0$ & $0$ & $0$ & $0$ & $0$ & $0$ & $0$ & $0.333$ & $0.667$ & $0.256$ & $0.077$ & $0.264$ & $0.659$ & $0.341$\\
\hline
$\tVE_{14}$ & $0$ & $0$ & $0$ & $0$ & $0$ & $0$ & $0$ & $0$ & $0$ & $0$ & $0$ & $0$ & $0$ & $0$ & $0$ & $0$ & $0.769$ & $0.231$ & $0.220$ & $0.549$ & $0.451$\\
\hline
$\tVE_{15}$ & $0$ & $0$ & $0$ & $0$ & $0$ & $0$ & $0$ & $0$ & $0$ & $0$ & $0$ & $0$ & $0$ & $0$ & $0$ & $0$ & $0.769$ & $0.231$ & $0.220$ & $0.549$ & $0.451$\\
\hline
$\tVE_{16}$ & $0$ & $0$ & $0$ & $0$ & $0$ & $0$ & $0$ & $0$ & $0$ & $0$ & $0$ & $0$ & $0$ & $0$ & $0$ & $0$ & $0$ & $0$ & $0.286$ & $0.714$ & $0.286$\\
\hline
$\tVE_{17}$ & $0$ & $0$ & $0$ & $0$ & $0$ & $0$ & $0$ & $0$ & $0$ & $0$ & $0$ & $0$ & $0$ & $0$ & $0$ & $0$ & $0$ & $0$ & $0.286$ & $0.714$ & $0.286$\\
\hline
$\tVE_{18}$ & $0$ & $0$ & $0$ & $0$ & $0$ & $0$ & $0$ & $0$ & $0$ & $0$ & $0$ & $0$ & $0$ & $0$ & $0$ & $0$ & $0$ & $0$ & $0$ & $0$ & $1$\\
\hline
$\tVE_{19}$ & $0$ & $0$ & $0$ & $0$ & $0$ & $0$ & $0$ & $0$ & $0$ & $0$ & $0$ & $0$ & $0$ & $0$ & $0$ & $0$ & $0$ & $0$ & $0$ & $0$ & $1$\\
\hline
$\tVE_{20}$ & $0$ & $0$ & $0$ & $0$ & $0$ & $0$ & $0$ & $0$ & $0$ & $0$ & $0$ & $0$ & $0$ & $0$ & $0$ & $0$ & $0$ & $0$ & $0$ & $0$ & $0$\\
\hline
$\tVE_{21}$ & $0$ & $0$ & $0$ & $0$ & $0$ & $0$ & $0$ & $0$ & $0$ & $0$ & $0$ & $0$ & $0$ & $0$ & $0$ & $0$ & $0$ & $0$ & $0$ & $0$ & $0$\\
\hline 
\end{tabular} 
%\end{tiny}
\caption{Probability matrix $\PM$: each element $\PME_{ij}$ gives the probability that transition $i$ is followed by transition $j$ after an arbitrary number of steps. This matrix is deduced from the adjacency matrix (Table~\ref{Tab:ex13L_AdjacencyMatrix}) by application of Eq.~(\ref{Eq:MatriceProba}).
}
\label{Tab:ex13L_ProbabilityMatrix}
\end{table}

The gated spectra obtained by applying the analytical formula (\ref{Eq:eS-intensity}) and (\ref{Eq:cS-intensity})
can be compared with the results obtained with a purely numerical approach.
We have performed the following procedure: 
\begin{enumerate}
\item List all possible cascades: starting from each transition, the different possible ways are determined thanks to the adjacency matrix, following a recursive algorithm.
\item Determine the probability associated with each cascade. 
For this, we first have to determine the probability that the deexcitation process
starts with the first transition of the cascade we are considering.
This is done by combining information contained in the emission probability vector and in the adjacency matrix.
For the rest of the cascade, each step is associated with a probability factor given by the corresponding adjacency-matrix element.
\item The gated probability $P_{\{G,i\}}$ for each transition $\tVE_i$ is obtained by summing the probabilities of all cascades that contain both $\tVE_i$ and the gates needed to pass the selection.
\end{enumerate}
The gated probability values we have obtained with the analytical and numerical approaches are strictly identical,
for all kinds of gate conditions. 
This can be seen in Figure~\ref{Fig:spectra_ex13L}, where gated spectra are presented with four different gate conditions.
The corresponding values of gated probabilities are listed in Table~\ref{Tab:ex13L_GatedProba}.

\begin{table}
\begin{tabular}{|c|c|c|c|c|}
\hline 
Transition $\tVE_i$ & $P_{\{G_1,i\}}$ & $P_{\{G_2,i\}}$ & $P_{\{G_3,i\}}$ & $P_{\{G_4,i\}}$ \\
\hline
$\tVE_1$ & $0.0258$ & $0.0207$ & $0.0432$ & $0.0673$ \\
\hline
$\tVE_2$ & $0.0525$ & $0.0421$ & $0.0880$ & $0.1370$ \\
\hline
$\tVE_3$ & $0.0127$ & $0.0047$ & $0.0047$ & $0.0177$ \\
\hline
$\tVE_4$ & $0.0670$ & $0.0593$ & $0.1290$ & $0.1907$ \\
\hline
$\tVE_5$ & $0.0086$ & $0.0076$ & $0.0076$ & $0.0166$ \\
\hline
$\tVE_6$ & $0.0119$ & $0$ & $0$ & $0.0119$ \\
\hline
$\tVE_7$ & $0.0124$ & $0$ & $0.0077$ & $0.0515$ \\
\hline
$\tVE_8$ & $0.0960$ & $0.0960$ & $0.1580$ & $0.2190$ \\
\hline
$\tVE_9$ & $0.0301$ & $0$ & $0.0077$ & $0.0382$ \\
\hline
$\tVE_{10}$ & $0$ & $0$ & $0$ & $0.0310$ \\
\hline
$\tVE_{11}$ & $0$ & $0$ & $0.0314$ & $0.0775$ \\
\hline
$\tVE_{12}$ & $0.1710$ & $0.0960$ & $0.1344$ & $0.2246$ \\
\hline
$\tVE_{13}$ & $0$ & $0$ & $0.0102$ & $0.0331$ \\
\hline
$\tVE_{14}$ & $0$ & $0$ & $0.0212$ & $0.0753$ \\
\hline
$\tVE_{15}$ & $0.0475$ & $0.0267$ & $0.0452$ & $0.0812$ \\
\hline
$\tVE_{16}$ & $0.1235$ & $0.0694$ & $0.0994$ & $0.1765$ \\
\hline
$\tVE_{17}$ & $0.0475$ & $0.0267$ & $0.0594$ & $0.1343$ \\
\hline
$\tVE_{18}$ & $0$ & $0$ & $0.0069$ & $0.0222$ \\
\hline
$\tVE_{19}$ & $0$ & $0$ & $0.0237$ & $0.0540$ \\
\hline
$\tVE_{20}$ & $0.1710$ & $0.0960$ & $0.1351$ & $0.2568$ \\
\hline
$\tVE_{21}$ & $0$ & $0$ & $0.0307$ & $0.0762$ \\
\hline
\end{tabular} 
\caption{Gated probabilities of the different transitions,
with the sets of gates used to obtain the spectra of Figure~\ref{Fig:spectra_ex13L}:
$G_1=\{\tVE_{12}\cdot \tVE_{20}\}$, 
$G_2=\{\tVE_{8}\cdot \tVE_{12}\cdot \tVE_{20}\}$, 
$G_3=\{\tVE_{4} + \tVE_{8} + \tVE_{12} + \tVE_{20}\}_3$, 
$G_4=\{\tVE_{4} + \tVE_{8} + \tVE_{12} + \tVE_{20}\}_2$.
}
\label{Tab:ex13L_GatedProba}
\end{table}

\subsection{Determination of the emission probability of a new transition}

Finally, we can check that our approach allows to determine the emission probability 
of a newly observed transition that is added to the level scheme.
This new transition is denoted by $\tVE_x$.
We make the following suppositions:
\begin{itemize}
\item $\tVE_x$ is directly connected to the previously known level scheme,
in the sense that at least its receiving level belongs to the known level list.
In this situation, all the cascades below $\tVE_x$ are already characterized.
\item Coincidence data allows to place correctly $\tVE_x$ in the level scheme.
\item The emission probability has been previously determined
for all other transitions present in the level scheme,
or, at least, all transitions involved in the relative intensity of $\tVE_x$ in the gated spectrum that is studied.
\end{itemize}
Then, is it possible to determine the emission probability $P_x$ of the new transition.
In practice, for the present study, 
we start from the nuclear structure described above 
(Figure~\ref{Fig:scheme_ex13L},
with the corresponding transition-space information given by Tables~\ref{Tab:ex13L_transitions} and \ref{Tab:ex13L_AdjacencyMatrix}),
and we pick a transition that will play the role of $\tVE_x$.
Information concerning $\tVE_x$ is deleted from the emission probability vector $\PV$ and adjacency matrix $\AM$.
It has to be recovered by using the information remaining in $\PV$ and $\AM$, 
together with the "observation" of a gated spectrum 
(here, this spectrum is previously calculated using the complete $\PV$ and $\AM$).
We are especially interested in recovering the value of $P_x$.

Let us first study the consequences of inserting a new transition $\tVE_x$ in the level scheme.
The transition-space dimension increases: 
the element $\tVE_x$ is added to the transition vector,
and the transition probability vector $\PV$ has to be completed with the corresponding value of $P_x$.
The impact on the adjacency matrix $\AM$ will be the following :
\begin{itemize}
\item The line and column corresponding to $\tVE_x$ have to be added.
\item For all transitions $\tVE_i$ arriving on the emitting level of $\tVE_x$,
the adjacency matrix elements $\AME_{ij}$ have to be re-normalized to take into account the new possible decay path:
the new values depend on $P_x$.
We have: $\AME_{ij}=P_j/\sum_{j^\prime}P_{j^\prime}$, where $\tVE_x$ is now part of the list of transitions $\tVE_{j^\prime}$.
\item The line added for $\tVE_x$ is similar to the lines corresponding to the transitions that have the same receiving level
as $\tVE_x$ (it is determined by the branching properties of the receiving level, unaffected by $P_x$).
\end{itemize}
The modification of $\AM$, in turn, has an impact on the probability matrix $\PM$,
and consequently on the gated probabilities given by Eq.~(\ref{Eq:cS-intensity}).
(We will only refer to the gated probabilities in combined spectra, 
since this formula allows to recover as a specific case 
the gated probabilities for elementary spectra given by Eq.~(\ref{Eq:eS-intensity}).)
Note that the $P_x$ value will affect the gated probability $P_{\{G,i\}}$ of a transition $\tVE_i$
if $\tVE_x$ is between $\tVE_i$ and one of the gates, or between one of the gates and $\tVE_i$, or between two gates.
Let us also specify how the gated probability $P_{\{G,x\}}$ of $\tVE_x$ depends on the emission probability $P_x$.
We notice that Eq.~(\ref{Eq:cS-intensity}) can be split in two terms, 
one (that will be denoted by $X$) treating the cases where $\tVE_x$ is above the sub-list of gates, 
and one (that will be denoted by $Y$) treating the cases where $\tVE_x$ is between two gates of the sub-list:
\be 
X(P_x)&=& 
\sum_{p=m}^N c_p(m)
\left[{
\sum_{\alpha=1}^{C^N_p}
\left[{
P_x \times 
\mP_{\tVE_x\ra g^{(\alpha)}_1} \times
\prod_{j=2}^{p} \mP_{g^{(\alpha)}_{j-1}\ra g^{(\alpha)}_j}
}\right]
}\right]\nn\\
Y(P_x)&=&
\sum_{p=m}^N c_p(m)
\left[{
\sum_{\alpha=1}^{C^N_p}
\left[{
P_{g_1^{(\alpha)}} \times
\sum_{h=1}^p 
\prod_{j=1}^p \mP_{\CME^h_{\alpha,j-1}\ra \CME^h_{\alpha,j}}
}\right]
}\right]\nn
\ee
$X$ and $Y$ both depend on $P_x$; 
however, the dependence of $X$ is explicit, since the probability matrix elements $\PME_{ij}$
that are involved do not depend on 
$P_x$ but only on the branching ratios of the levels below $\tVE_x$.
We can then write $X(P_x) = P_x \times C_x$, 
where $C_x$ is a "constant" that does not depend on $P_x$, 
and can be calculated using only the reduced transition-space information.
On the other hand, $Y$ is expressed in terms of probability matrix elements
that depend on $P_x$, through the modified adjacency values occuring within the cascade.

Let us now turn to the study of the relative gated intensity $I^{(r)}_x(G)$,
that can be measured in a gated spectrum with respect to a reference transition $\tVE_{ref}$:
\be 
I^{(r)}_x(G)&=&\frac{N_{\{G,x\}}}{N_{\{G,ref\}}} = \frac{P_{\{G,x\}}}{P_{\{G,ref\}}}\nn
\ee
The gated probabilities $P_{\{G,x\}}$ and $P_{\{G,ref\}}$ are given by Eq.~(\ref{Eq:cS-intensity}).
In general, these values are affected by $P_x$ and the relative gated intensity 
has the following dependence:
\be 
I^{(r)}_x(G)&=& \frac{P_x\times C_x + Y(P_x)}{{P_{\{G,ref\}}}(P_x)}\nn
\ee
This can be solved for $P_x$ using a numerical iterative procedure:
\begin{enumerate}
\item Propose a value of $P_x$ in the emission probability vector $\PV$.
\item Deduce the modifications that have to be done in the adjacency matrix.
\item Re-calculate the probability matrix according to Eq.~(\ref{Eq:MatriceProba}).
\item Apply Eq.~(\ref{Eq:cS-intensity}) to determine the new gated probabilities.
\item Compare the ratio ${P_{\{G,x\}}}/{P_{\{G,ref\}}}$ 
to the relative intensity $I^{(r)}_x(G)$ measured in the gated spectrum.
\item Modify the value of $P_x$ for a new iteration, until convergence is reached. 
\end{enumerate}
We have applied this procedure to our example, using a simple dichotomy.
Trantition $\tVE_{16}$ was used as the reference transition,
and the different gate conditions shown in Figure~\ref{Fig:spectra_ex13L} have been applied.
All the transitions different from $\tVE_{ref}$ and from the gates 
have been treated in turn as being the new transition $\tVE_x$.
In all cases, the correct value of $P_x$ was recovered, except in one case: the last transition, $\tVE_{21}$.
Indeed, for this last transition, changing the value of $P_x$ has no impact on the gated probabilities,
since no adjacency element is affected, and $\tVE_{21}$ is never on top of a selected cascade. 
Actually, changing $P_{\tVE_{21}}$ only means changing the primary feeding of the emitting level $(B2)_1$, 
which cannot be reflected by a gated spectrum since $\tVE_{21}$ is a final transition.
For such a transition, the emission probability has to be determined without gate conditions.
Let us note, however, that it is not a disadvantage for our purpose:
indeed, our focus is on the transitions situated in the high region of the level scheme, 
where multi-gating is needed to make observations. 
Lack of knowledge about transition probabilities (such as $P_{\tVE_{21}}$) 
that do not affect such spectra has, by definition, no consequence.

Let us now turn to the case of transitions situated in the higher part of the level scheme.
More specifically, we consider that the new transition $\tVE_x$ is situated above the set of gates, 
and above the reference transition $\tVE_{ref}$.
In this case, $P_{\{G,ref\}}$ has no dependence on $P_x$,
while $P_{\{G,x\}}$ reduces to $P_x \times C_x$.
Since $P_{\{G,ref\}}$ and $C_x$ can be calculated using only the reduced transition-space information,
we can determine $P_x$ directly once $I^{(r)}_x(G)$ is measured in the gated spectrum:
\be
P_x &=& \frac{I^{(r)}_x(G)\times P_{\{G,ref\}}}{C_x}\nn
\ee
We have applied this procedure to our example,
still using trantition $\tVE_{16}$ as the reference transition,
and applying the different gate conditions shown in Figure~\ref{Fig:spectra_ex13L}.
The correct value of $P_x$ was obtained for all the transitions situated above the set of gates.

\section{Summary and outlook}
\label{Sec:SumOutlook}

In the present work, 
we have addressed the issue of recovering the absolute probability of a transition
through the measurement of intensities appearing in multi-gated spectra,
using different kinds of gate conditions (explicit or optional).
We have presented the base of a formalism
that allows to treat this problem following an analytic approach,
and we have demonstrated formulas linking the gated probability of a gamma ray
with two objects that characterize the transition space of the excited nucleus:
the emission probability vector $\PV$,
and the probability matrix $\PM$. 
The former is linked to the primary feeding of the levels, and branching ratios;
the latter, whose elements $\PME_{ij}$ give the probability that a transition $\tVE_j$ occurs
after a transition $\tVE_i$ has taken place (whatever the number of steps inbetween),
is deduced from the transition adjacency matrix $\AM$
by the analytic formula presented by Demand et al.~\cite{Demand11}.
We have found the graph-theory framework used in this reference to be very fructful 
and promizing for the type of problems to be addressed in gamma spectroscopy.
Although the intensity problem we address can in principle be treated in a purely numerical way,
the analytic approach allows to gain more control on the complexity of the analysis,
and offers both a way to check the results and a powerful tool to extract emission probabilities
in the case of new transitions on top of the set of gates.

Although the basic principles are soundly set down in this article,
some developments are needed before the present formalism 
can be applied to extract emission probabilities from real experimental data.
We will address in future work several generalizations,
concerning both the physics of the deexciting nucleus and the characteristics of the experimental setup.
Concerning the nucleus, we should include the following possibilities:
\begin{itemize}
\item different deexcitation modes (e.g. electronic conversion, eletron-positron pair emission);
\item cases where the deexcitation cascade is cut before reaching the ground-state: presence of isomeric states,
or nuclear disintegration occurring from an excited state;
\item existence of degenerate transitions, which has an impact on the nature of the gating condition: namely, 
if a gate energy corresponds to the energy of several transitions, an option ("or") in the coincidence condition is introduced.
\end{itemize}
Note that, in order to treat the last point, the structure of the gate condition will have to be generalized
beyond the two cases defined in this article (explicit/optional).
It is anyway a useful development to consider gate conditions with a mixed explicit and optional structure,
depending on the part of the level scheme where the gates are situated.
Concerning now the experimental setup, we should specify the following properties:
\begin{itemize}
\item response of the detector system: detection efficiency, eventually including the role of angular correlations between the emitted gamma rays;
\item possibility of having experimental data filtered by a multiplicity threshold;
\item treatment of the back-ground.
\end{itemize}

These improvements will allow one to apply the analytic formalism introduced here in the case of real data.
They could eventually lead to the development of a dedicated piece of software.
This approach stresses the importance to obtain accurate (rather than approximately estimated) values of the emission probabilities, 
even when they concern weak gamma rays that can only be accessed via multi-gated spectra.
Precise results can be used for instance as a criterion to check the tentative placement of a new transition in the level scheme:
consistent values have to be obtained when using different gate conditions.
Most importantly, the emission probabilities contain fundamental information that should be used 
to improve our knowledge of nuclear structure and reactions.

\appendix

\section{Examples of optional conditions and associated spectra}
\label{App:spectrum-development-examples}

An optional condition is denoted by $\mG\{g_1+...+g_N\}_m$;
it involves a list of $N$ optional gates $L=\{g_1,...,g_N\}$, 
and $m$ is the minimal number of open gates among this list.
The combined set of events associated with this condition is $\mE(\mG)$; 
it is represented by the combined spectrum $\cS(\mG)$.
We want to develop $\cS(\mG)$ in terms of positive elementary spectra 
$S_\alpha=S(G_\alpha=\{g^{(\alpha)}_1\cdot ... \cdot g^{(\alpha)}_p\})$.
Each positive elementary condition $G_\alpha$
involves a list of $p$ gates $L^{(\alpha)}=\{g^{(\alpha)}_1,...,g^{(\alpha)}_p\}$
extracted from $L$, with $m\leq p \leq N$.
%These elementary spectra represent sets of events 
%associated with explicit conditions that impose $p$ open gates, 
%where $L^{(\alpha)}=\{g^{(\alpha)}_1,...,g^{(\alpha)}_p\}$ is a sublist of $L$ and $n\leq p \leq N$.
In this appendix, we illustrate with specific examples the formulas that are derived in the text and summarized below:
\begin{itemize}
\item Tiling relation (\ref{Eq:tiling}) to express the combined set as non-overlapping exclusive elementary sets:
\be 
\mE(\mG=L_{m/N})&=& \bigcup_{n=m}^N \bigcup_{\beta=1}^{C^N_n} E(G_\beta(n,N-n,L))\nn
\ee
\item The resulting relation (\ref{Eq:combined-spectrum}) between combined and exclusive elementary spectra:
\be 
\cS(\mG) &=& \sum_{n=m}^{N} \sum_{\beta=1}^{C^{N}_n} \eS(G_\beta(n,N-n,L))
= \sum_{n=m}^{N} \sigma(n,N-n,L)\nn
\ee
\item Expression of exclusive sum-spectra in terms of positive sum-spectra. We remind that a sum-spectrum is the summation of all the elementary spectra of the same order, 
that can be defined from the same gate list $L$. 
This expression is given by Eq. (\ref{Eq:sum-spectrum-relation-2}):
\be 
\sigma (n,N-n,L) &=& \sum_{p=n}^N a_{n,p} \; \sigma(p,L)
\quad \rm{with} \; a_{n,p}=(-1)^{p-n} C^p_n
\nn
\ee
This leads to the final expression (\ref{Eq:combined-spectrum-development}) of the combined spectrum:
\be 
\cS(\mG) &=& \sum_{p=m}^{N} c_p(m)\;\sigma(p,L) 
\quad \rm{with} \; c_m(p)=\sum_{n=m}^p a_{n,p}=\sum_{n=m}^p (-1)^{p-n} C^p_n
\nn
\ee
\item Concerning the spiked spectrum associated with condition $\mG$, 
we also give the expression of $\mE(\mG)$ as the union of positive elementary sets, 
according to Eq. (\ref{Eq:simple-union}),
and the corresponding biased sum-spectrum
given by Eq. (\ref{Eq:spiked-spectrum}):
\be 
\mE(\mG)=\mE(L_{m/N}) &=& \bigcup_{\alpha=1}^{C^N_m} E(G_\alpha(m,L))\nn\\
\cSs(\mG)&=&\sigma(m,L)\nn
\ee
\end{itemize}

\subsection{Optional gate condition $\mG=\{g_1+g_2+g_3\}_2$}

In this example, $N=3$, $m=2$, and $L=\{g_1,g_2,g_3\}$.
\begin{itemize}
\item Combined set expressed by the tiling relation:
\be 
\mE(\{g_1+g_2+g_3\}_2)
&=& E_{g_1g_2\bar{g}_3} \cup E_{g_1g_3\bar{g}_2} \cup E_{g_2g_3\bar{g}_1} \cup E_{g_1g_2g_3} \nn
\ee
\item Combined spectrum expressed as a sum of exclusive elementary spectra:
\be 
\cS(\{g_1+g_2+g_3\}_2)
&=& \eS_{g_1g_2\bar{g}_3} + \eS_{g_1g_3\bar{g}_2} + \eS_{g_2g_3\bar{g}_1} + \eS_{g_1g_2g_3} 
=\sigma(2,1,L)+\sigma(3,0,L)
\nn
\ee
where the exclusive sum-spectrum of order $(2,1)$ is:
\be 
\sigma(2,1,L)=\sigma(2,1,\{g_1,g_2,g_3\})
&=& \eS_{g_1g_2\bar{g}_3} + \eS_{g_1g_3\bar{g}_2} + \eS_{g_2g_3\bar{g}_1} \nn
\ee
and the exclusive sum-spectrum of order $(3,0)$, equivalent to the positive sum-spectrum of order $3$, is:
\be 
\sigma(3,0,L)=\sigma(3,0,\{g_1,g_2,g_3\}) &=& \sigma(3,\{g_1,g_2,g_3\}) = \eS_{g_1g_2g_3} \nn
\ee
\item Expression of the exclusive sum-spectrum $\sigma(2,1,L)$ 
in terms of positive sum-spectra, as given by Eq. (\ref{Eq:sum-spectrum-relation-2}):
\be 
\sigma(2,1,L) &=& a_{2,2}\sigma(2,L)+a_{2,3}\sigma(3,L)
=\sigma(2,L)-3\sigma(3,L)\nn
\ee
with
\be 
\sigma(2,L) &=& \eS_{g_1g_2} + \eS_{g_1g_3} + \eS_{g_2g_3}\nn\\
\sigma(3,L) &=& \eS_{g_1g_2g_3}\nn
\ee
As a result:
\be 
\cS(\{g_1+g_2+g_3\}_2) &=& \left[\sigma(2,L)-3\sigma(3,L)\right]+\sigma(3,L)
= \sigma(2,L)-2\sigma(3,L)\nn\\
&=& \eS_{g_1g_2} + \eS_{g_1g_3} + \eS_{g_2g_3} - 2 \eS_{g_1g_2g_3}\nn
\ee
Note that for such a reduced list of gates, the final result can easily be obtained 
in a pedestrian approach, applying intuitively the development of exclusive spectra 
involving one closed gate:
\be 
\eS_{g_1g_2\bar{g}_3} &=& \eS_{g_1g_2} - \eS_{g_1g_2g_3} = g_1g_2(1-g_3) \nn\\
\eS_{g_1g_3\bar{g}_2} &=& \eS_{g_1g_3} - \eS_{g_1g_2g_3} = g_1g_3(1-g_2) \nn\\
\eS_{g_2g_3\bar{g}_1} &=& \eS_{g_2g_3} - \eS_{g_1g_2g_3} = g_2g_3(1-g_1) \nn
\ee
which are specific examples of the general relation (\ref{Eq:spectrum-develop-step2a}).
We recover the final result:
\be 
\cS(\{g_1+g_2+g_3\}_2) &=& \eS_{g_1g_2} + \eS_{g_1g_3} + \eS_{g_2g_3} - 2 \eS_{g_1g_2g_3}\nn
\ee
\item We finally consider the spiked spectrum. 
The combined set can be expressed as the union of all positive elementary sets of order $m=2$:
\be 
\mE(\mG)&=& \bigcup_{\alpha=1}^{C^3_2} E(G_\alpha(2,L)) = E_{g_1g_2} \cup E_{g_1g_3} \cup E_{g_2g_3} \nn
\ee
The corresponding summation of elementary spectra 
(which involves multi-counting of events in the overlapping regions of the united sets) 
gives the spiked spectrum:
\be 
\cSs(\mG)&=& \sigma(2,L) = \sum_{\alpha=1}^{C^3_2} S(G_\alpha(2,L)) 
= \eS_{g_1g_2} + \eS_{g_1g_3} + \eS_{g_2g_3} \nn
\ee
The relation between combined and spiked spectra is:
\be 
\cS(\mG)&=& \cSs(\mG) - 2 \eS_{g_1g_2g_3}\nn
\ee
which, again, can be found intuitively in this simple example:
one can see directly that the events of the overlapping part $E_{g_1g_2g_3}$
are counted three times in the spiked spectrum,
since they belong to all three sets $E_{g_1g_2}$, $E_{g_1g_3}$ and $E_{g_2g_3}$.
\end{itemize}

\subsection{Gate condition $G=\{g_1+g_2+g_3+g_4\}_2$}

In this example, $N=4$, $m=2$, and $L=\{g_1,g_2,g_3,g_4\}$.
With only one more gate in the optional list, one finds that the pedestrian approach
to express the combined spectrum in terms of positive elementary spectra
is already much more tedious, 
and the analytic expressions that have been derived are now helpful.
\begin{itemize}
\item Combined set expressed by the tiling relation:
\be 
\mE(\{g_1+g_2+g_3+g_4\}_2)
&=& \bigcup_{n=2}^4 \bigcup_{\beta=1}^{C^4_n} E(G_\beta(n,4-n,L)) \nn\\
&=& \bigcup_{\beta=1}^{C^4_2} E(G_\beta(2,2,L))
\bigcup_{\beta^\prime=1}^{C^4_3} E(G_{\beta^\prime}(3,1,L)) 
\bigcup_{\beta^{\prime\prime}=1}^{C^4_4} E(G_{\beta^{\prime\prime}}(4,0,L))\nn
\ee
where
\be 
\bigcup_{\beta=1}^{C^4_2} E(G_\beta(2,2,L)) &=&
E_{g_1g_2\bar{g}_3\bar{g}_4}
\cup E_{g_1g_3\bar{g}_2\bar{g}_4}
\cup E_{g_1g_4\bar{g}_2\bar{g}_3}
\cup E_{g_2g_3\bar{g}_1\bar{g}_4}
\cup E_{g_2g_4\bar{g}_1\bar{g}_3}
\cup E_{g_3g_4\bar{g}_1\bar{g}_2} \nn\\
\bigcup_{\beta^\prime=1}^{C^4_3} E(G_{\beta^\prime}(3,1,L)) &=&
E_{g_1g_2g_3\bar{g}_4}
\cup E_{g_1g_2g_4\bar{g}_3}
\cup E_{g_1g_3g_4\bar{g}_2}
\cup E_{g_2g_3g_4\bar{g}_1} \nn\\
\bigcup_{\beta^{\prime\prime}=1}^{C^4_4} &=& E_{g_1g_2g_3g_4} \nn
\ee
\item Combined spectrum expressed as a sum of exclusive elementary spectra:
\be 
\cS(\{g_1+g_2+g_3+g_4\}_2)
&=& \sum_{n=2}^4 \sum_{\beta=1}^{C^4_n} \eS(G_\beta(n,4-n,L)) \nn\\
&=& \sum_{\beta=1}^{C^4_2} \eS(G_\beta(2,2,L)) 
+ \sum_{\beta^\prime=1}^{C^4_3} \eS(G_{\beta^\prime}(3,1,L))
+ \sum_{\beta^{\prime\prime}=1}^{C^4_4}\eS(G_{\beta^{\prime\prime}}(4,0,L))\nn\\
&=& \sigma(2,2,L) + \sigma(3,1,L) + \sigma(4,0,L)\nn
\ee
where the sum-spectra $\sigma$ are:
\be 
\sigma(2,2,L) &=&
\eS_{g_1g_2\bar{g}_3\bar{g}_4}
+\eS_{g_1g_3\bar{g}_2\bar{g}_4}
+\eS_{g_1g_4\bar{g}_2\bar{g}_3}
+\eS_{g_2g_3\bar{g}_1\bar{g}_4}
+\eS_{g_2g_4\bar{g}_1\bar{g}_3}
+\eS_{g_3g_4\bar{g}_1\bar{g}_2} \nn\\
\sigma(3,1,L) &=&
\eS_{g_1g_2g_3\bar{g}_4}
+\eS_{g_1g_2g_4\bar{g}_3}
+\eS_{g_1g_3g_4\bar{g}_2}
+\eS_{g_2g_3g_4\bar{g}_1} \nn\\
\sigma(4,0,L) &=& \eS_{g_1g_2g_3g_4} \nn
\ee
\item Expression of the exclusive sum-spectra $\sigma(n,N-n,L)$ 
in terms of positive sum-spectra $\sigma(p,L)$.
Following the pedestrian approach, each term can be developped recursively 
according to:
\be 
\eS_{g_1g_2\bar{g}_3\bar{g}_4}
&=& \eS_{g_1g_2\bar{g}_3} - \eS_{g_1g_2g_4\bar{g}_3} \nn\\
&=& (\eS_{g_1g_2}-\eS_{g_1g_2g_3}) - (\eS_{g_1g_2g_4}-\eS_{g_1g_2g_3g_4})\nn
\ee
In the end, we recover the result expressed by the analytic formula:
\be 
\cS(\mG=\{g_1+g_2+g_3+g_4\}_2) &=& \sum_{p=2}^{4} c_p(m)\,\sigma(p,L)\nn
%= \sum_{p=2}^{4} \left[\sum_{n=2}^p a_{n,p}\right]\,\sigma(p,L) \nn
\ee 
Term $p=2$:
\be 
c_p(m)=\sum_{n=2}^p a_{n,p} = \sum_{n=2}^p (-1)^{p-n}C^p_n &=& 1 \nn\\
\sigma(2,L)&=& \eS_{g_1g_2} + \eS_{g_1g_3} + \eS_{g_1g_4} + \eS_{g_2g_3} + \eS_{g_2g_4} + \eS_{g_3g_4} \nn
\ee
Term $p=3$:
\be 
c_p(m)=\sum_{n=2}^p a_{n,p} = \sum_{n=2}^p (-1)^{p-n}C^p_n &=& -3 + 1 = -2 \nn\\
\sigma(3,L)&=& \eS_{g_1g_2g_3} + \eS_{g_1g_2g_4} + \eS_{g_1g_3g_4} + \eS_{g_2g_3g_4} \nn
\ee
Term $p=4$:
\be 
c_p(m)=\sum_{n=2}^p a_{n,p} = \sum_{n=2}^p (-1)^{p-n}C^p_n &=& 6 -4 +1 = 3 \nn\\
\sigma(4,L)&=& \eS_{g_1g_2g_3g_4}\nn
\ee
The combined spectrum is then expressed as:
\be 
\cS(\mG) &=& \sigma(2,L) - 2\times\sigma(3,L) + 3\times\sigma(4,L) \nn
\ee
Note that the coefficients $c_p(m)=\sum_{n=m}^p (-1)^{p-n}C^p_n$ 
can be obtained by column summation in a universal table
that contains the coefficients $a_{n,p}$ (Table~\ref{Tab:coefs}).
\item Let us finally consider the spiked spectrum. 
The union of all positive elementary sets of order $m=2$ is now:
\be 
\mE(\mG)&=& \bigcup_{\alpha=1}^{C^4_2} E(G_\alpha(2,L)) 
= E_{g_1g_2} \cup E_{g_1g_3} \cup E_{g_1g_4} \cup E_{g_2g_3} \cup E_{g_2g_4} \cup E_{g_3g_4} \nn
\ee
The corresponding summation of elementary spectra 
(which involves multi-counting of events in the overlapping region of the united sets) 
gives the spiked spectrum:
\be 
\cSs(\mG)&=& \sigma(2,L) = \sum_{\alpha=1}^{C^4_2} S(G_\alpha(2,L)) = 
\eS_{g_1g_2} + \eS_{g_1g_3} + \eS_{g_1g_4} + \eS_{g_2g_3} + \eS_{g_2g_4} + \eS_{g_3g_4} \nn
\ee
The relation between combined and spiked spectra is:
\be 
\cS(\mG)&=& \cSs(\mG) - 2\times\sigma(3,L) + 3\times\sigma(4,L) \nn
\ee
\end{itemize}

\end{document}